\documentclass[structabstract]{aa} 
\usepackage{graphicx}
\usepackage{txfonts}
\usepackage{appendix}
\usepackage{natbib}
\bibpunct{(}{)}{;}{a}{}{,}
\def\nbody{$N$-body}
\begin{document}
   \title{Disc-protoplanet interaction}

   \subtitle{Influence of circumprimary radiative discs on self-gravitating protoplanetary bodies in binary star systems}

   \author{M. Gyergyovits
    \fnmsep\thanks{Send any requests to the first author}
          \inst{1}
          \and 
          S. Eggl \inst{1,2}
          \and
          E. Pilat - Lohinger \inst{1,4}
          \and
          Ch. Theis \inst{3}
          }

   \institute{Institute for Astrophysics (IfA), University of Vienna,
              T\"urkenschanzstrasse 17, A-1180 Vienna\\
              \email{markus.gyergyovits@univie.ac.at}
          \and
             IMCCE - Observatoire de Paris,
             77 Avenue Denfert-Rochereau, F-75014 Paris
          \and
             Planetarium Mannheim, 
             Wilhelm-Varnholt-Allee 1, 68165 Mannheim
          \and
             Department of Physics, University of Graz,
             Universit\"{a}tsplatz 5, 8010 Graz   
        }

   \date{}

  \abstract
   {More than 60 planets have been discovered so far in systems that harbour two stars, some of which have binary semi-major axes as small as 20 au. It is well known that the formation of planets in such systems is strongly influenced by the stellar components, since the protoplanetary disc and the particles within are exposed to the gravitational influence of the binary. However, the question on how self-gravitating protoplanetary bodies affect the evolution of a radiative, circumprimary disc is still open.}
   {We present our 2D hydrodynamical GPU-CPU code and study the interaction of several thousands of self-gravitating particles with a viscous and radiative circumprimary disc within a binary star system. To our knowledge this program is the only one at the moment that is capable to handle this many particles and to calculate their influence on each other and on the disc.}
   {We performed hydrodynamical simulations of a circumstellar disc assuming the binary system to be coplanar. Our grid-based staggered mesh code relies on ideas from ZEUS-2D, where we implemented the FARGO algorithm and an additional energy equation for the radiative cooling according to opacity tables. To treat particle motion we used a parallelised version of the precise Bulirsch - Stoer algorithm. Four models in total  where computed taking into account (i) only \nbody $ $ interaction, (ii) \nbody $ $ and disc interaction, (iii) the influence of computational parameters (especially smoothing) on \nbody $ $ interaction, and (iv) the influence of a quiet low-eccentricity disc while running model (ii). The impact velocities where measured at two different time intervals and were compared.}
   {We show that the combination of disc- and \nbody $ $ self-gravity can have a significant influence on the orbit evolution of roughly Moon sized protoplanets.}
   {Not only gas drag can alter the orbit of particles, but the gravitational influence of the disc can accomplish this as well. The results depend strongly on the state of the disc (i.e. quiet or dynamically evolving) - according to encounter-probability distributions, planet formation can be strongly altered if there is a dynamically evolving gas disc -  and  also on the  smoothing parameter.}

   \keywords{Accretion, accretion disks -- Hydrodynamics -- Radiative transfer -- Methods: numerical -- Planets and satellites: formation --  binaries (including multiple): close --
               }
  
   \maketitle
%
%________________________________________________________________
%__________________________________________________________________
\section{Introduction}
\label{sec:introduction}
 Most main-sequence stars in the solar neighbourhood are members of multiple systems \citep{1991A&A...248..485D}, and the distribution of their separations peaks at about 50 au \citep{eggleton-2006}. This suggests a significant influence of the secondary star on the circumprimary protoplanetary disc. Planets have been discovered in mid-separation binaries, with semi-major axes of about 20 au, for example, Gl86 \citep{2000A&A...361..265S,2006A&A...459..955L}, HD 41004 \citep{2004A&A...426..695Z}, $\gamma$ Cephei \citep{2003ApJ...599.1383H,2007A&A...462..777N}, and the recently discovered HD 196885 \citep{2011A&A...528A...8C} and $\alpha$ Centauri B \citep{2012Natur.491..207D}, which has led to a growing interest in understanding planetary formation processes in general.
In mid-separation binary star systems, the secondary and the heavily truncated and possibly distorted disc causes the formation and evolution of planets throughout several stages of the planet-forming process. The protoplanetary disc is truncated by the companion star through gravitational interaction, as was shown by  \citet{1994ApJ...421..651A} and \citet{1994MNRAS.268...13S}. This shortens the viscous lifetime of the disc and  consequently limits the period in which gas-giants can form. During accretion phases to km-sized particles, the second star can force the shape and orientation of planetesimal orbits, which can stop planetary formation \citep[e.g.][and references therein]{2011CeMDA.111...29T}.\\
Despite the considerable progress this field has seen over the past decades  \citep{2000ApJ...543..328M,2002A&A...396..219B,2004RMxAC..22...99L,2004A&A...427.1097T,2006Icar..183..193T,2007ApJ...660..807Q,2007ApJ...666..436H,2008ApJ...679.1582G,2011CeMDA.111...29T}, many open questions remain, especially with regard to accretion stages where km-sized particles are expected to grow into planetary embryos within a gaseous disc.
This process demands fast runaway and oligarchic growth phases, which in turn require low encounter and impact velocities ($dv$)  between protoplanetary bodies. Previous studies \citep{2000ApJ...543..328M,2004A&A...427.1097T,2006Icar..183..193T} have shown that coupling the protoplanetary bodies to gas via gas drag creates a size-dependent phasing of protoplanetary orbits that decreases $dv$ for equal-sized objects but increases it for all other types of impacts \citep{2011CeMDA.111...29T}.  Introducing small inclinations between the static, axisymmteric, circumprimary gas disc and the binary orbital plane might facilitate accretion \citep{2009ApJ...698.2066X,2010ApJ...708.1566X}. Recently, \citet{2013Natur.493..381K} showed that wide binaries can also influence the evolution of embedded planetary systems even Gyr after their formation. These studies are based on either pure \nbody $ $ simulations that neglect the influence of a gas disc \citep{2000ApJ...543..328M,2002A&A...396..219B,2004RMxAC..22...99L,2007ApJ...660..807Q,2007ApJ...666..436H,2008ApJ...679.1582G}  
 or only take gas drag caused by a static circular disc \citep{2004A&A...427.1097T,2006Icar..183..193T,2011CeMDA.111...29T} into account. A static circular disc, however, is not realistic for close binary star systems since the secondary exerts spiral density waves within the disc, which will presumably alter the evolution of protoplanets by either drag force or gravitational force. In an eccentric binary system even the assumption of circularity does not hold any more. This was shown by \citet{2008A&A...486..617K} and \citet{2008A&A...487..671K}.\\

The influence of an evolving gas-disc on protoplanetary accretion has been studied first by \citet{2008MNRAS.386..973P} in two dimensions and more recently by \citet{2011A&A...528A..40F} in 3D. Both studies used a locally isothermal disc as underlying model without taking the self-gravity of particles into account. \citet{2012A&A...539A..18M} showed, that radiative discs have far lower eccentricities than their locally isothermal counterparts. This means that, changing the radiative properties of the protoplanetary disc can lead to a dynamically quiet environment more suitable for planet formation than the local isothermal approach.  

While most of the studies described in the previous paragraph focused on embryo formation via protoplanetary accretion, little attention has been given so far to the penultimate stage in the formation scenario, where relatively large protoplanets interact gravitationally with each other as well as with a dynamically evolving disc. The lack of such investigations into this regime might be related to the huge computational requirements for simulations where self-gravitation of the protoplanets and the gravitational interaction with the disc is more important than gas drag.

We here introduce a GPU-CPU 2D hydrodynamic grid code that combines hydrodynamic radiative disc computations with highly accurate \nbody $ $ simulations. The development of this new code allows us to investigate the interplay between a circumprimary gas disc and the orbital evolution of embedded self-gravitating protoplanets on time-scales of one hundred binary periods. The application of our code using various models shows the differences for planetary formation, when taking into account \\

\begin{tabular}[h]{ll}
  (i) & binary - disc interactions \\
  (ii)& binary - protoplanet interaction \\
  (iii) & binary - protoplanet - disc interactions. \\ 
 & \\
\end{tabular}

In section~\ref{sec:code_description} we describe important features of our code and present results for standard tests in section~\ref{sec:test} . The results from our numerical investigations of combined disc-particle systems are presented in section~\ref{sec:particle_disc}. A summary of our findings and concluding remarks are given in section~\ref{sec:sum_con}.

%__________________________________________________________________
%__________________________________________________________________

\section{Code description}
\label{sec:code_description}

Our code is based on the ZEUS-2D solver for hydrodynamical equations on a two-dimensional polar grid \citep{1992ApJS...80..753S}. It was originally developed by \citet{1999A&A...347..442K} for simulations  of galaxies and computes interactions of up to three different fluids in one simulation \citep{2004A&A...418..959T}. 
The heart of the code consists of a staggered-mesh finite-difference method using operator splitting and a first-order integrator in time to handle the source step. The advective terms are solved by a second-order upwind algorithm in time - for a detailed description see \citet{1992ApJS...80..753S}. The adaptation of the code to hydrodynamical simulations of gaseous discs in tight binary systems required the following major modifications: (i) an accelerated coordinate system that is centred on the primary star; (ii) implementing of the so-called FARGO acceleration algorithm \citep{2000A&AS..141..165M} to speed up calculations; (iii) we partly rewrote the code to run on NVIDIA graphics cards in double-precision accuracy to run simulations that include the full gravitational interaction of several thousand particles with the disc. \\
The equations of motion of the hydrodynamic part of the code in an accelerated coordinate system are given by
 \begin{equation}
  \begin{array}{l} 
  \displaystyle  \partial _t v_{r} + v_{r} \partial _r v_{r} +  \frac{v_{\phi}}{r}\partial _{\phi} v_{r} -  \frac{v_{\phi}^{2}}{r} = -\frac{1}{\Sigma} \partial _r P +  \bigtriangledown \cdot \tens{T}_r - \partial _r \Psi + f_{r}\\
   \\
  \displaystyle \partial _t v_{\phi} + v_{r} \partial _r v_{\phi} +  \frac{v_{\phi}}{r} \partial _{\phi} v_{\phi} + \frac{v_{\phi} v_{r}}{r} = -\frac{1}{\Sigma r} \partial _{\phi} P + \bigtriangledown \cdot \tens{T}_{\phi} - \frac{1}{r} \partial _{\phi} \Psi + f_{\phi},
  \end{array}
  \label{eq_of_motion}
 \end{equation}
where $v_{r}$, $v_{\phi}$ are the velocity components in \textit{r} and $\phi$ direction. \textit{P} denotes to the pressure, $\Sigma (r,\phi)$ is the vertically integrated density, $\Psi(\vec{r})$ is the gravitational potential exerted by all bodies (a detailed description is given below in this section) and $f_{r}$, $f_{\phi}$ denote the accelerations due to the non-inertial frame. Here, the secondary as well as the massive particles moving in the disc will contribute to this acceleration. The effects of viscosity are treated as discussed in \citet{1999MNRAS.303..696K} and \citet{1959flme.book.....L}. We implemented the standard viscous stress tensor, but used  the turbulent viscosity coefficient $\nu _t$ instead of the molecular viscosity. This leads to the equations
 \begin{equation}
   \bigtriangledown \cdot \tens{T}_r = \frac{1}{r} \left[ \partial _r \left( rT_{rr} \right) + \partial _{\phi} T_{r \phi} \right] - \frac{T_{\phi \phi}}{r}
  \label{divtensr}
 \end{equation}
 and
 \begin{equation}
 \bigtriangledown \cdot \tens{T}_{\phi} =   \partial _r \left( T_{r \phi } \right) + \partial _{\phi} \left( \frac{T_{\phi \phi}}{r} \right) + 2 \frac{T_{r \phi}}{r},
  \label{divtensphi}
 \end{equation}
 where $T_{rr}$, $T_{\phi \phi}$, $T_{r \phi}$ are the components of the viscous stress tensor:
  \begin{equation}
  \begin{array}{l} 
   T_{rr} = 2\eta \partial _r v_r + \left(\xi -\frac{2}{3} \eta \right) \bigtriangledown \cdot \vec{v} \\
   \\
   T_{\phi \phi} = 2\eta \left(\frac{1}{r} \partial _{\phi} v_{\phi} + \frac{v_r}{r}\right) + \left(\xi -\frac{2}{3} \eta \right) \bigtriangledown \cdot \vec{v}\\
   \\
   T_{r \phi} = \eta \left(\partial _r v_{\phi} + \frac{1}{r} \partial _{\phi} v_r -\frac{v_{\phi}}{r} \right).\\
  \end{array}
  \label{compstresstens}
 \end{equation}
 $\bigtriangledown \cdot \vec{v}$ denotes the divergence of the velocity and $\eta$ and $\xi$ are the shear and bulk viscosity, respectively. In our simulations the bulk viscosity was set to zero. For the shear viscosity we used $\eta = \Sigma \nu _t$. Depending on the simulation scenario, the turbulent viscosity coefficient was either set constant or according to the $\alpha$ prescription by \citet{1973A&A....24..337S}. We found that better results - fewer post-shock oscillations in the Sod shock-tube test \citep{1978JCoPh..27....1S} - can be achieved when we applied an artificial viscosity of von Neumann \& Richtmyer type \citep{1992ApJS...80..753S}. \\
The gravitational potential exerted on the disc by the masses ($M_i$) is given by
\begin{equation}
 \label{eqpotdisc}
 \Psi(\vec{r})=-\frac{GM_1}{\mid \vec{r} - \vec{r}_1\mid}-\frac{GM_2}{\mid \vec{r} - \vec{r}_2\mid} - \sum_{i=3}^n \frac{GM_i}{\mid \vec{r} - \vec{r}_i + \epsilon \mid},
\end{equation}
where $\vec{r}_1$, $\vec{r}_2$ are the position vectors and $M_1$, $M_2$ are the masses of the two stars (primary and secondary). The secondary is moving outside the computational grid. Furthermore, G denotes the gravitational constant, $\vec{r}_i$, $M_i$ are the position vectors and masses of the smaller bodies, and $\vec{r}$ is the point of interest in the disc. 
For bodies moving in the grid, gravity softening was used via the $\epsilon$ softening parameter \citep[equation 3]{2006MNRAS.370..529D}. \\
For closure conditions, the code offers (i) the widely known locally isothermal approach, (ii) a globally isothermal model, and (iii) a radiative disc model. 
To solve the required energy equation we followed \citet{2012A&A...539A..18M} with one small exception: 
instead of mixing adiabatic and isothermal sound speed, we used only the latter one. For reasons of numerical stability we also used a lowest value for the optical depth. 
The vertically integrated energy equation reads
\begin{equation}
 \partial _t e + \bigtriangledown \cdot \left(e \vec{v} \right) = -p \bigtriangledown \cdot \vec{v} + Q_{+} - Q_{-} - 2 H \bigtriangledown \cdot \vec{F},
 \label{enquation}
 \end{equation}
where $e$ is the internal energy density, $Q_{+}$ the heating source term, $Q_{-}$ the cooling source term \citep{2012A&A...539A..18M}. 
The radiative transport in the plane of the disc is treated in the flux-limited diffusion approximation. The flux is given by
\begin{equation}
 \vec{F} = -\frac{\lambda c 4 a T^3}{\rho \kappa} \bigtriangledown T,
\end{equation}
where $c$ denotes the speed of light, $a$ the radiation constant, $\rho$ the mid-plane density, $\kappa$ the Rosseland mean opacity, and $\lambda$ is the flux-limiter \citet[eqn. 9]{1989A&A...208...98K} and \citet{2008A&A...487L...9K}.
The initial radial velocity $v_r$ was set to zero, and $v_{\phi}$ is sub-Keplerian, meaning that the pressure terms as well as the viscosity are taken into account when the initial values for $v_{\phi}$ are computed from equation (\ref{eq_of_motion}). \\
To handle the \nbody $ $ calculations accurately we implemented a  Bulirsch - Stoer integrator \citep{bulirsch-stoer-1964} parallelised to run on the GPU and thus gaining a speed factor of up to 40.

%__________________________________________________________________
\section{Application of the code}
 \subsection{Models}
 \label{sec:particle_disc}
Because the tests showed (see section \ref{sec:test}) that our code provides satisfactory performance, we applied our new code to study systems with full gravitational interaction between the binary stars, the disc, and the protoplanets. 
In close binary star systems such as $\gamma$ Cephei - which we took as our model - it can be assumed that close encounters between protoplanets and additional interaction with frequently occuring spiral waves induced by the secondary star in the gas disc lead to momentum and energy exchange of such an amount that this might increase the semi-major axis and eccentricity of the protoplanets to high values and thus hinder agglomeration and merging of particles and protoplanets. \\ \citet{2012A&A...539A..18M} showed that in such systems the mass-weighted disc eccentricity and longitude of disc pericenter of the gas disc can change their dynamical behaviour from damped oscillation to a constant small one (for the first) or from rotation to oscillation (the second). Such changes in dynamical evolution of the disc might also affect movement of protoplanets and may either render planet formation impossible or provide an environment where agglomeration is not strongly affected by the disc. \\ \\
Therefore we investigated 
\begin{itemize}
\item whether the combined gravitational influence (stars, protoplantes and disc) dampens or increases the orbital forcing of the protoplanets,  and
\item to which degree the choice of the smoothing parameter affects these simulations.
\end{itemize}
For this purpose we ran the following simulations:
\begin{description}
\item[\textbf{reference model:}] here we included only the binary-disc interaction.
\item[\textbf{model a1:}] several thousand self-gravitating protoplanets distort the disc gravitationally but no back-reaction from the disc on the particles is considered. 
At the same time, the disc and the bodies move under the gravitational influence of the binary star.
\item[\textbf{model a2:}] the same initial conditions as model a1 are used except for a higher smoothing parameter when calculating the particle - particle forces. 
\item[\textbf{model b1:}] the full gravitational interaction between particles, the disc and the binary star is taken into account.
\item[\textbf{model b2:}] we recalculated the reference model for 50000 yr, reset disc mass to its initial value, and then inserted particles in the disc taking into account the full gravitational interaction between particles, the disc and the binary star.
\end{description}
The results of models a1, a2, b1 and b2 were compared with the disc evolution of our reference model and with each other.

%__________________________________________________________________
\subsection{Numerical setup}
 \label{sec:numsetup}

For the simulations we used a two-dimensional logarithmic polar grid ranging from 0.5 au to 8 au, and the initial conditions given in Table \ref{table1}. The masses of the primary and secondary were 1.4 $M_{\sun}$ and 0.4 $M_{\sun}$. Their semi-major axis was 20 au and the chosen eccentricity was 0.4. These values are close to those of $\gamma$ Cephei provided by \citet{2007A&A...462..777N} and were also used in the study of \citet{2012A&A...539A..18M}. Our initial disc mass was 0.01 $M_{\sun}$ and we used an alpha viscosity prescription, where $\alpha$ has a value of $5\times 10^{-3}$. The adiabatic index $\gamma$ was set to $7/5$. The initial density and temperature profiles of the gas disc follow a $\propto r^{-1}$ distribution. Our initial disc aspect ratio (H/r) was 0.05, which is a commonly used value. Moreover, we chose different grid resolutions in $\phi$ and $r$ directions to allow for locally square cells where $\Delta r \approx r \Delta \phi$. We ran simulations with 2048 self-gravitating protoplanetary bodies and the two stellar components of the binary, where we fixed the masses to values that would follow from a complete coagulation of dust - the dust-to-gas-ratio being $10^{-2}$. This gives particles of mass $\sim 0.016$ $M_{\oplus}$. As a first approximation we removed the term connected to the radiative flux in the disc in the energy equation, since its contribution to the whole equation is several orders of magnitude lower than that of the other terms. The total integration time for our reference model as well as the simulation runs of models a1, a2, b1, and b2 was 100 binary periods, corresponding to 6670 years. Initially, the protoplanets were distributed on 32 circular orbits centred on the binary's primary star. On each of the 32 orbits, 64 particles were evenly distributed so that all particles are started on linearly spaced circular orbits ranging from 1 to $\sim4$ au in radial direction and chosen in such a way that $\Delta r \approx r_0 \Delta \phi$, where $r_0 = 1$ au. Thus the last ring is close to the critical semi-major axis $a = 3.920 \pm 0.009$ au where stable motion is still possible according to the studies by \citet{1999AJ....117..621H} and \citet{2002CeMDA..82..143P}. We chose this quite artificial approach of an ordered state because we did not wish to introduce additional perturbation in the system. We assumed that due to the given gravitational influences the system will adjust itself during the simulation. Of course, one could have tested a variety of different initial conditions and their influence on the outcome of the simulations, but this would be a task of its own and will be presented in a forthcoming article. 
\begin{table}
\caption{Initial conditions for the simulations}
\label{table1}
\centering
\begin{tabular}{l l}
\hline \hline                            %begin tabular
 Primary mass ($M_{\mathrm{1}}$)					& 1.4 $M_{\sun}$					\\
 Secondary mass ($M_{\mathrm{2}}$)					& 0.4 $M_{\sun}$					\\
 Semi-major axis ($a_{\mathrm{bin}}$)				& 20 au								\\
 Eccentricity ($e_{\mathrm{bin}}$)					& 0.4								\\
\hline
 Disc mass ($M_d$)									& 0.01 $M_{\sun}$					\\
 Viscosity ($\alpha$)								& $5\times 10^{-3}$					\\
 Adiabatic index ($\gamma$)							& 7/5								\\
 Mean-molecular weight ($\mu$)						& 2.35								\\
 Initial density profile ($\Sigma$)					& $\propto r^{-1}$					\\ 
 Initial temperature profile (T)					& $\propto r^{-1}$					\\
 Initial disc aspect ratio (H/r)					& 0.05								\\
 Grid ($N_{\mathrm{r}} \times N_{\mathrm{\phi}}$)	& 254 $\times$ 576					\\
 Computational domain ($r_{\mathrm{min}}$ - $r_{\mathrm{max}}$)	& 0.5 - 8 au			\\ 	
\hline
 Protoplanet mass ($M_{\mathrm{p}}$)					& $\sim 0.016$ $M_{\oplus}$	\\
 Number of protoplanets ($N_{\mathrm{p}}$)				& 2048								\\
\hline                                   %end tabular
\end{tabular}      
\end{table}

Since these particles are moving within the hydrodynamical grid, we used a gravity softening of $\epsilon = 0.6 H_p$, a value commonly found in literature. We did not include the more sophisticated treatment suggested by \citet{2012A&A...541A.123M}. For the chosen particle masses the corresponding softening sphere is slightly larger than the particles' Roche lobe with respect to the primary. Using gravity softening for the particle - particle interactions as well ($\epsilon_p = 10^{-5}$ au), we decided not to take protoplanet merging into account. For protoplanets with mean lunar density, the associated softening distance is of the order of their respective radii. Such a weak softening is not expected to influence the computations significantly, but it prevents very small computational time-steps. For this reason processes such as gravitational focusing of the bodies or coagulation of particles are probably not inhibited during the simulation. Given the current choices of initial and boundary conditions, the gas drag exerted by the disc on the protoplanets would be two to three orders of magnitude lower than the self-gravity of the two protoplanets at opposing outer edges of the truncated disc. We therefore neglected possible influences of gas drag and concentrated only on gravitational particle - particle and particle - disc interactions. To compare all our results we kept the secondary moving on a fixed ellipse according to its initial conditions. Reflecting boundary conditions at the inner boundary and zero gradient boundary condition at the outer boundary were applied for all simulations (for a short description on boundary condition details see also \citet{2012A&A...539A..18M}). Modelling the inner boundary is quite a delicate task since we do not have any information about this region. A zero-gradient boundary condition will create a (eccentric) hole in the vicinity of this boundary \citep{2008A&A...486..617K}. Taking a non-reflecting boundary condition \citep{1996MNRAS.282.1107G}, one also has to make assumptions about disc behaviour inside the boundary (which means outside the computational domain). We have chosen this rigid boundary since this initial density distribution needs to adjust itself to ongoing perturbation caused by the secondary and therefore will move inward at first. Of course, this boundary will reflect every wave into the computational regime and cause artificial superpositions of waves. But since the density is very high at the inner regions we expect these reflected waves to be damped very fast. The initial density of the gas disc was modelled according to the following power-law distribution:

 \begin{equation}
 \Sigma (r) = \Sigma_{0} r^{-n} f(r)$, $ \quad n = 1,
 \end{equation}
 where $\Sigma_{0}$ is the value of the vertical averaged density at r = 1. To avoid unrealistically low gas-densities we applied a floor value ($fl$), which is given by $fl = \Sigma_{0} \times 10^{-7}$. The density was modified by the function $f(r)$, which emulates a truncated disc.  For this purpose we introduced the Fermi function
 \begin{equation}
  f(r) = \frac{1}{1+e^{\delta (r - r_ {t})}}, \qquad \delta = \frac{\ln (10^{7}/r-1)}{r_{\mathrm{max}}-r_ {t}}
 \end{equation}
with $r_{t} = 5$ au.
The radial velocity ($v_r$) of the disc was set to zero and the azimuthal velocity ($v_{\phi}$) was adjusted taking the pressure gradient and the viscosity of the disc into account.
To analyse the disc behaviour and compare our work with others we also computed the mass-weighted eccentricity  ($e_{\mathrm{mw}}$) and argument of periastron of the gas disc ($\varpi_{\mathrm{mw}}$) where we followed \citet{2012A&A...539A..18M}, and which are given by
\begin{equation}
 e_{mw}= \left[\int_{r_{\mathrm{min}}}^{r_{\mathrm{max}}} \int_{0}^{2 \pi}  \Sigma e r dr d\phi \right] \times  \left[\int_{r_{\mathrm{min}}}^{r_{\mathrm{max}}} \int_{0}^{2 \pi}  \Sigma r dr d\phi  \right]^{-1}
\end{equation}
\begin{equation}
 \varpi_{mw} = \left[\int_{r_{\mathrm{min}}}^{r_{\mathrm{max}}} \int_{0}^{2 \pi}  \Sigma \varpi r dr d\phi \right] \times  \left[\int_{r_{\mathrm{min}}}^{r_{\mathrm{max}}} \int_{0}^{2 \pi}  \Sigma r dr d\phi  \right]^{-1}
\end{equation}
and e and $\varpi$ are the eccentricity and argument of pericenter computed for each cell separately. The integrals where evaluated by summation over all grid cells. This definition favours the inner parts of the disc, which are more dense, and thus represents the evolution of the region - at least in these parameters - in which we are interested in.

%__________________________________________________________________
\section{Results and discussion}
 \label{sec:results}
 \subsection{Evolution of the disc}
\begin{figure}[h]
 \resizebox{\hsize}{!}{\begin{tabular}{l}
                          \includegraphics{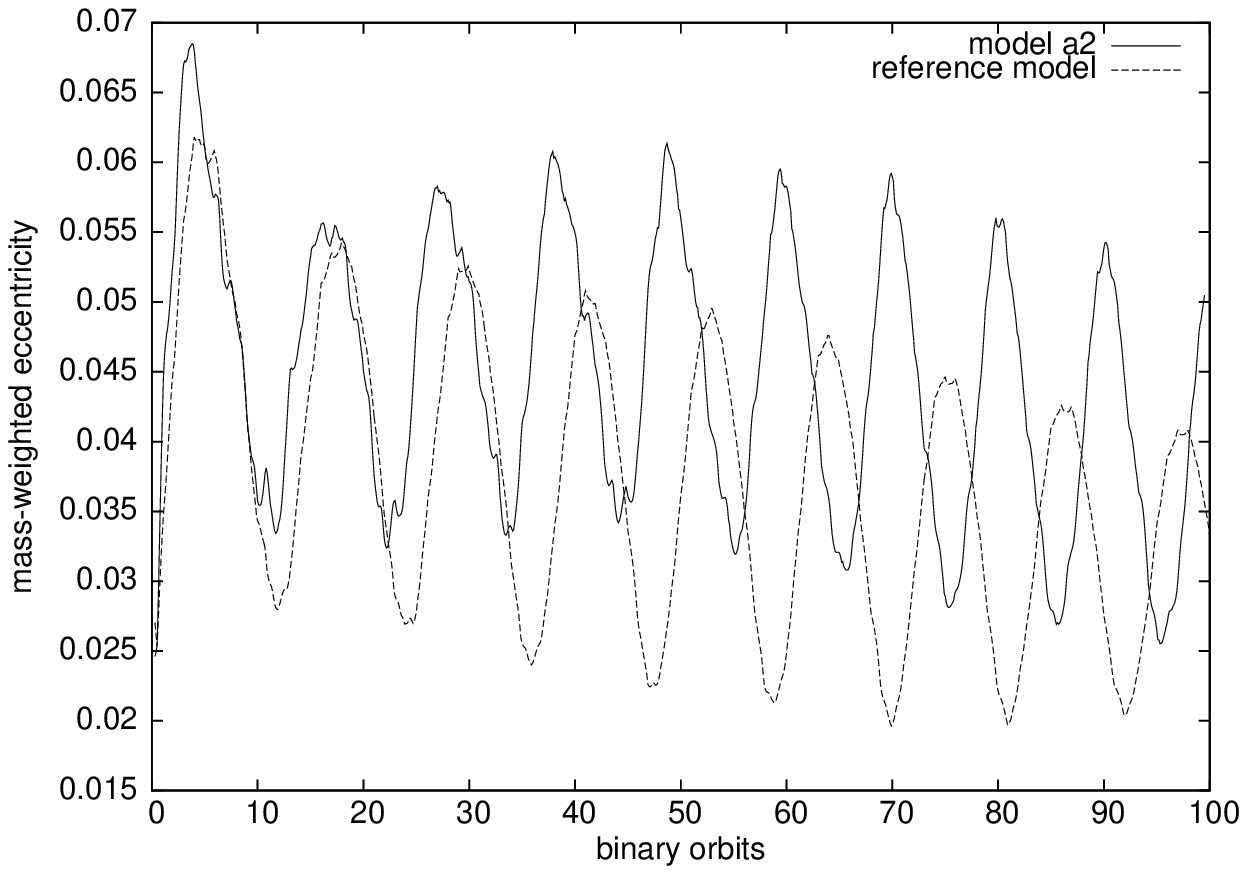} \\
                          \includegraphics{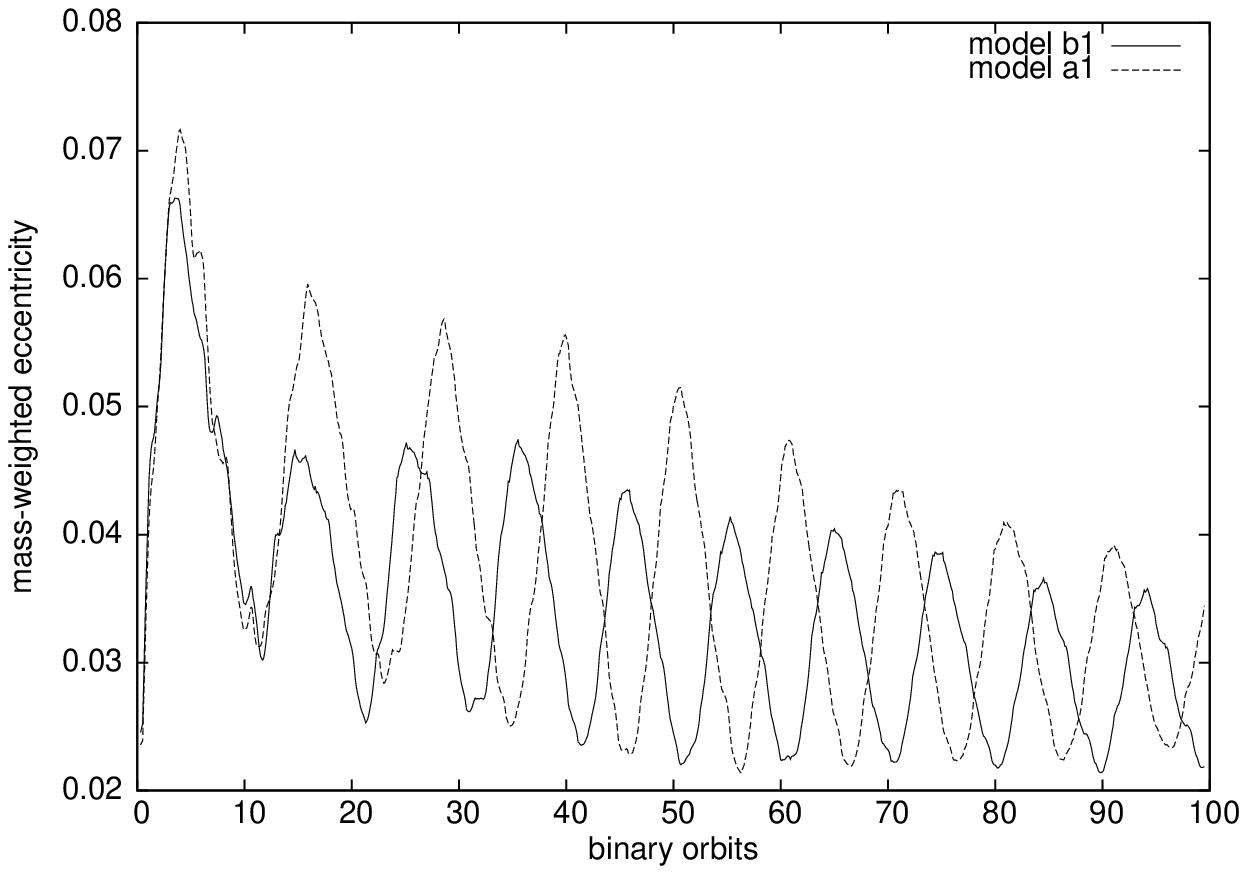}
                       \end{tabular}
                      }
 
 \caption{Time evolution of the mass-weighted eccentricity for a disc without particles (reference model) and model a2 (upper plot), and models a1 and b1 (lower plot). Time is given in terms of binary orbits, where one orbit corresponds to 66.7 yr. See \ref{sec:particle_disc} for details.}
   \label{figscompdiscs2}%
\end{figure}
\begin{figure}
 \resizebox{\hsize}{!}{\begin{tabular}{l}
                          \includegraphics{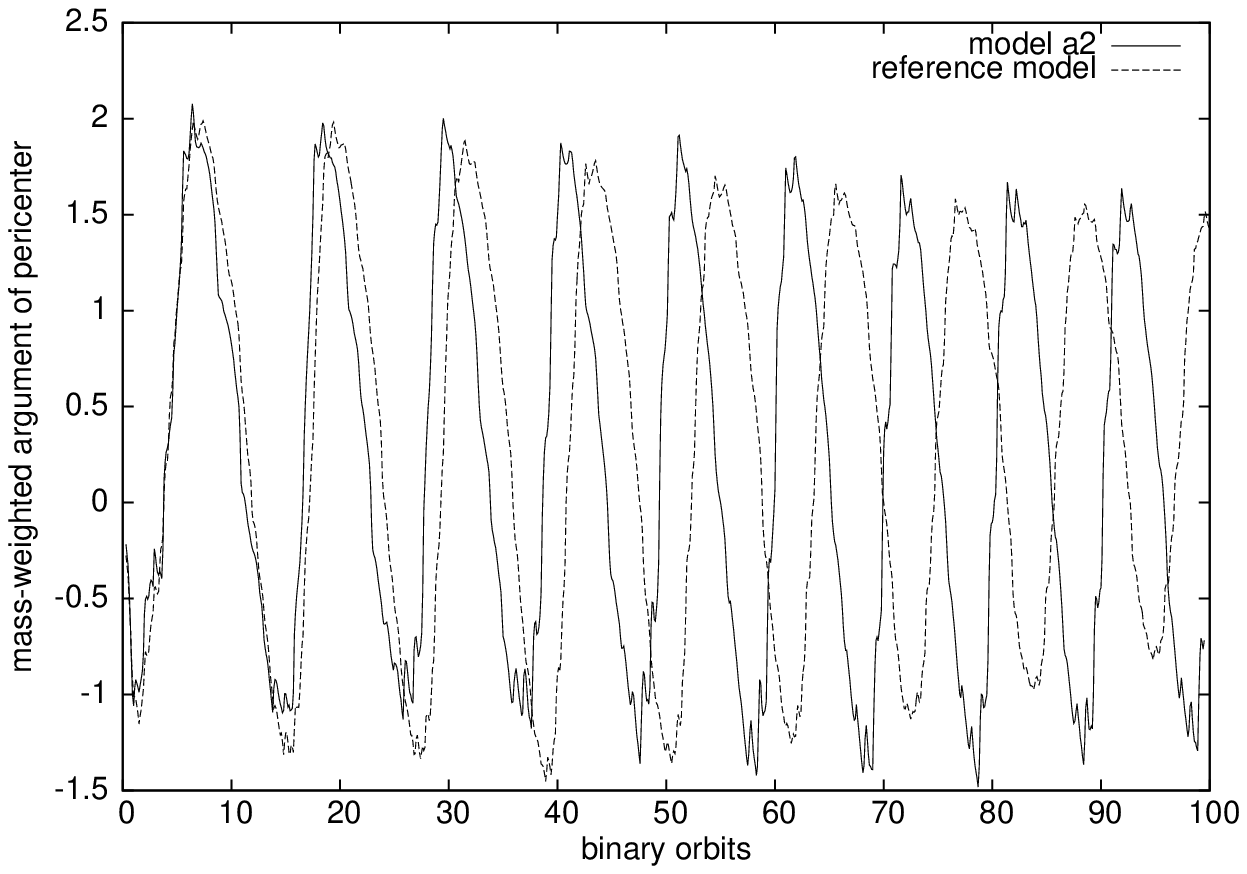} \\
                          \includegraphics{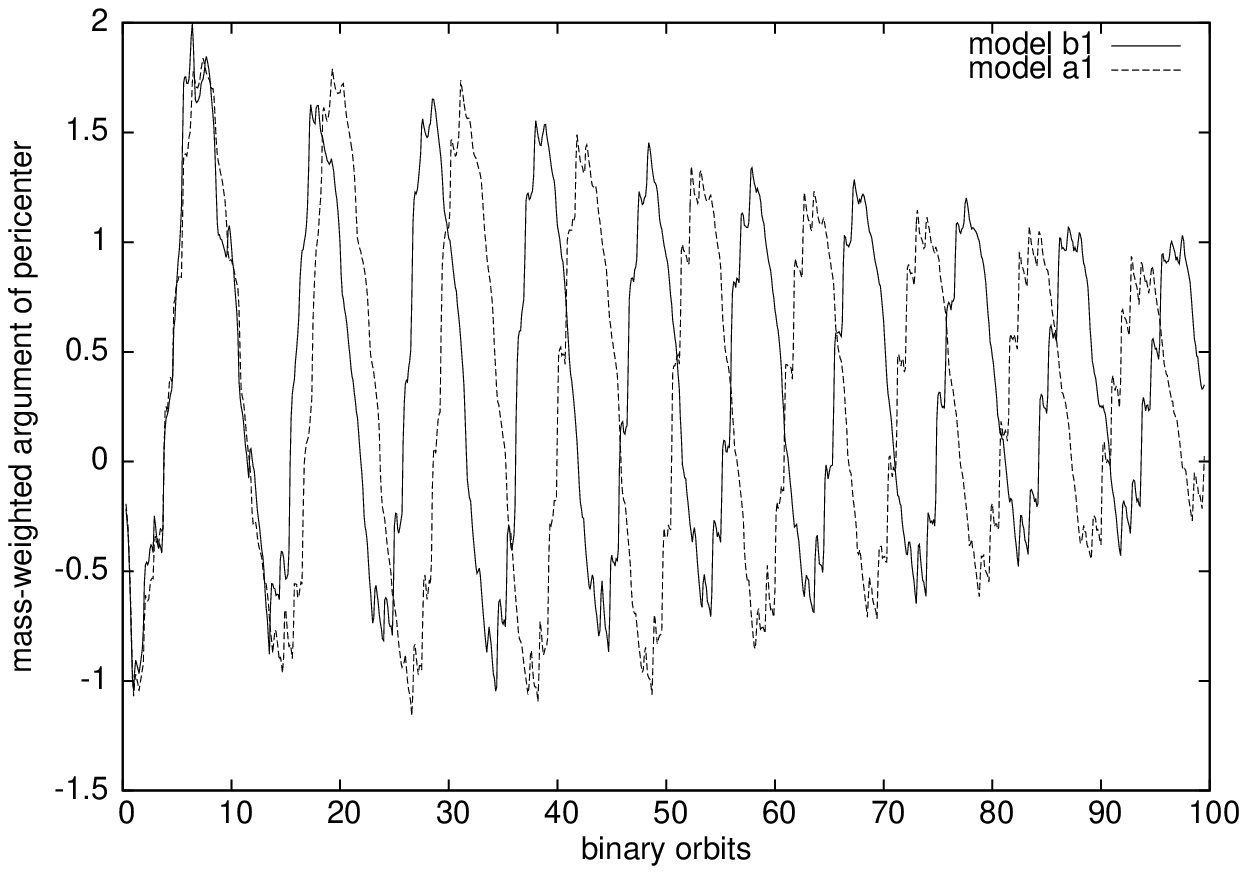}
                       \end{tabular}
                      }
 \caption{Time evolution of the mass-weighted argument of the pericenter for a disc without particles (reference model) and model a2 (upper plot), and models a1 and b1 (lower plot). Time is given in terms of binary orbits, where one orbit corresponds to 66.7 yr. }
   \label{figscompdiscs1}%
\end{figure}
\begin{figure}
 \resizebox{\hsize}{!}{\begin{tabular}{l}
                          \includegraphics{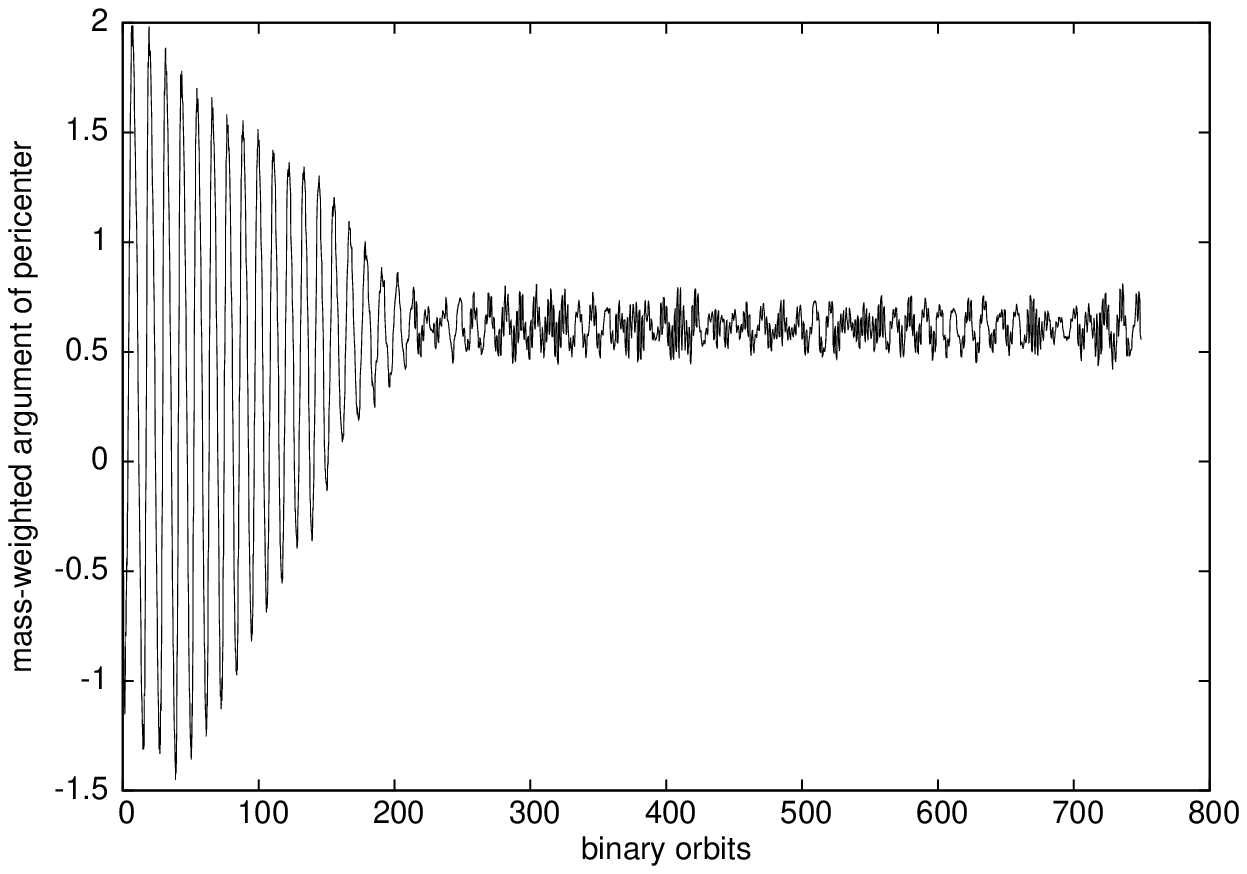} \\
                          \includegraphics{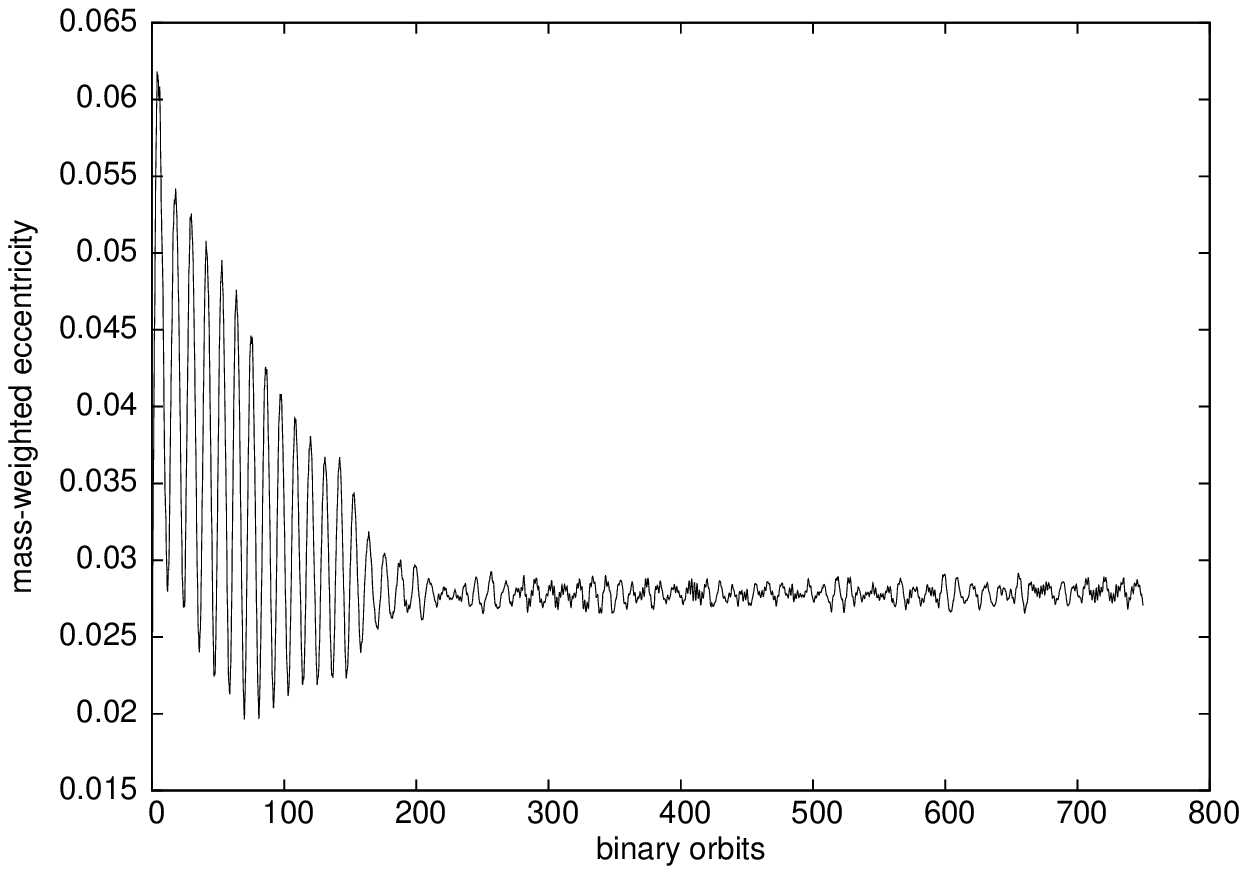}
                       \end{tabular}
                      }
 \caption{Time evolution of mass-weighted argument of the pericenter for a disc without particles (upper plot) and of mass-weighted eccentricity (lower plot) for 50000 yr after applying a running window average. Time is given in terms of binary orbits, where each orbit corresponds to 66.7 yr. We find transition from circulation to oscillation within 200 binary orbits for $\varpi_{mw}$ and a subsequent oscillation around $\approx$ 0.6 rad. A damped oscillation is visible for $e_{mw}$, reaching a value around $\approx$ 0.0275. }
   \label{refmod50000yrs}%
\end{figure}

In Figures \ref{figscompdiscs2}, \ref{figscompdiscs1}, and \ref{refmod50000yrs} we show the evolution of the mass-weighted eccentricity and mass-weighted argument of the pericenter for all models. To eliminate small oscillation caused by the periodical approach of the secondary (orbital period $\sim$ 66.7 yr) we applied a running window filter. For all of the models the eccentricity of the disc, which initially was zero, was increased to values of nearly 0.06-0.07 and reached $\sim$ 0.03 (model a1, b1) or $\sim$ 0.035 (reference model and model a2) at the end of the simulation (Figure \ref{figscompdiscs2}). The argument of the pericenter is rotating retrograde at the beginning of the simulation, but as time proceeds, the circulation turns into a damped oscillation around 0.5 radians for models a1 and b1 (Figure \ref{figscompdiscs1}, lower plot). For the reference model and model a2 (Figure \ref{figscompdiscs1}, upper plot) this effect seems to be present as well, but not as pronounced as in the first case. The results for the reference model agree well with those obtained by \citet{2012A&A...539A..18M}. \\
The extended integration of our reference model (Figure \ref{refmod50000yrs}) shows a similar behaviour as models a1 and b1.  We find transition from circulation to oscillation within 200 binary orbits for $\varpi_{mw}$ and a subsequent oscillation around $\approx$ 0.6 rad. A damped oscillation is visible for $e_{mw}$, reaching a value of about $\approx$ 0.0275. A similar transitional behaviour was also found in \citet{2012A&A...539A..18M}. \\
A comparison of Figures \ref{figscompdiscs2} and \ref{figscompdiscs1} indicates that a protoplanetary disc seems to damp the evolution of $e_{mw}$ and $\varpi_{mw}$ such that $e_{mw}$ for models a1 and b1 is lower than $e_{mw}$ for the reference model and the amplitude of oscillation of $\varpi_{mw}$ is also decreased. On the other hand, we see that an increasing  protoplanetary gravitational softening parameter for \nbody $ $ interaction increases the amplitudes for $e_{mw}$ as well as for $\varpi_{mw}$, indicating that this parameter has a significant influence on the computation.  
\\
Using an initially truncated disc not only depicts the response of the disc due to a perturber in a better way, it also diminishes the mass loss. This is because using a simple $r^{-1}$ density distribution places much material outside the Hill sphere of the primary where it is strongly affected by the perturbations of the secondary. This material is quickly removed from the disc within a few revolutions of the second star. This behaviour was found by the authors during test runs and can also be seen in Figure 12 of \citet{2012A&A...539A..18M} (small jump within the first 1000 years).  The mass loss during our simulations with respect to the initial disc mass was 0.68\% for the reference model,  0.52\% for model a1, 0.75\% for model a2,  0.42\% for model b1, and 0.08\% for model b2.\\
 Since we are using a rigid boundary condition at the inner boundary we can not provide data about mass loss at this boundary.  For all models we use an alpha value of $\alpha = 0.005$.

%__________________________________________________________________________________________________

\subsection{Evolution of particles}
\begin{figure}
 \resizebox{\hsize}{!}{\includegraphics{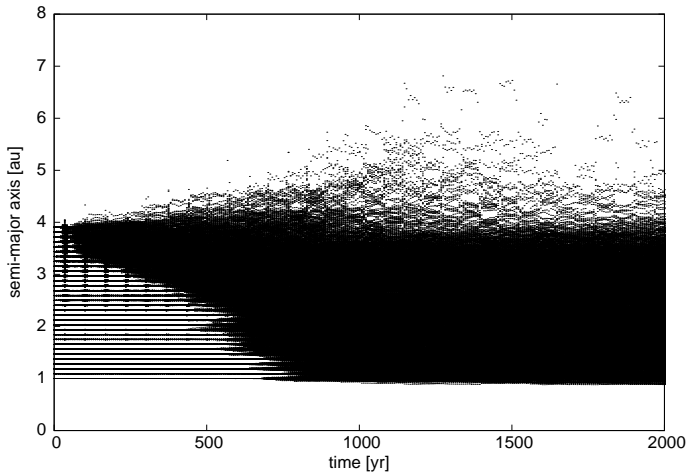}}
 \caption{Time evolution of the particle semi-major axis influenced by the disc (model b1). 
We find a transition from ordered motion close to the initial positions on the grid (left lower corner) to a distribution of semi-major axes that spreads across the whole stable region of the disc within 900 yr.}
   \label{figscompdisca}%
\end{figure}

Because periodical perturbations of the secondary alter the disc, the embedded particle orbits are affected by the changes in the gas disc and the direct perturbations of the secondary. The strong perturbations experienced by the outer disc propagate inwards through particle-particle interactions that cause strong deviations from the protoplanets' initial orbits. This process takes $\sim$ 900 years, as shown for model b1 in Figure \ref{figscompdisca}. We find a similar behaviour for model a1 with an increased evolution time of $\sim$2200 years, for model a2 where the time amounts to $\sim$ 3500, and for model b2 it takes $\sim$ 1100 years (all not shown).
\begin{figure}
 \resizebox{\hsize}{!}{\begin{tabular}{l}
                          \includegraphics{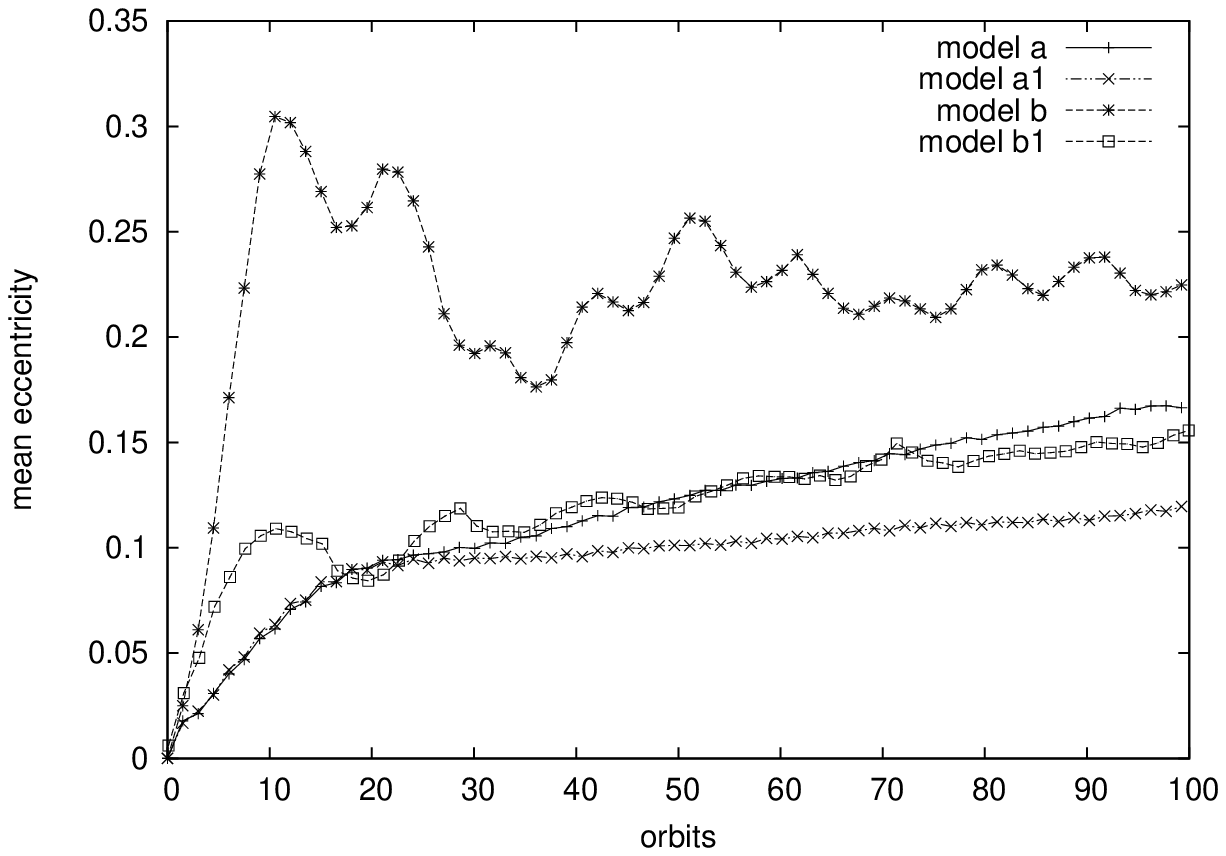}\\
                          \includegraphics{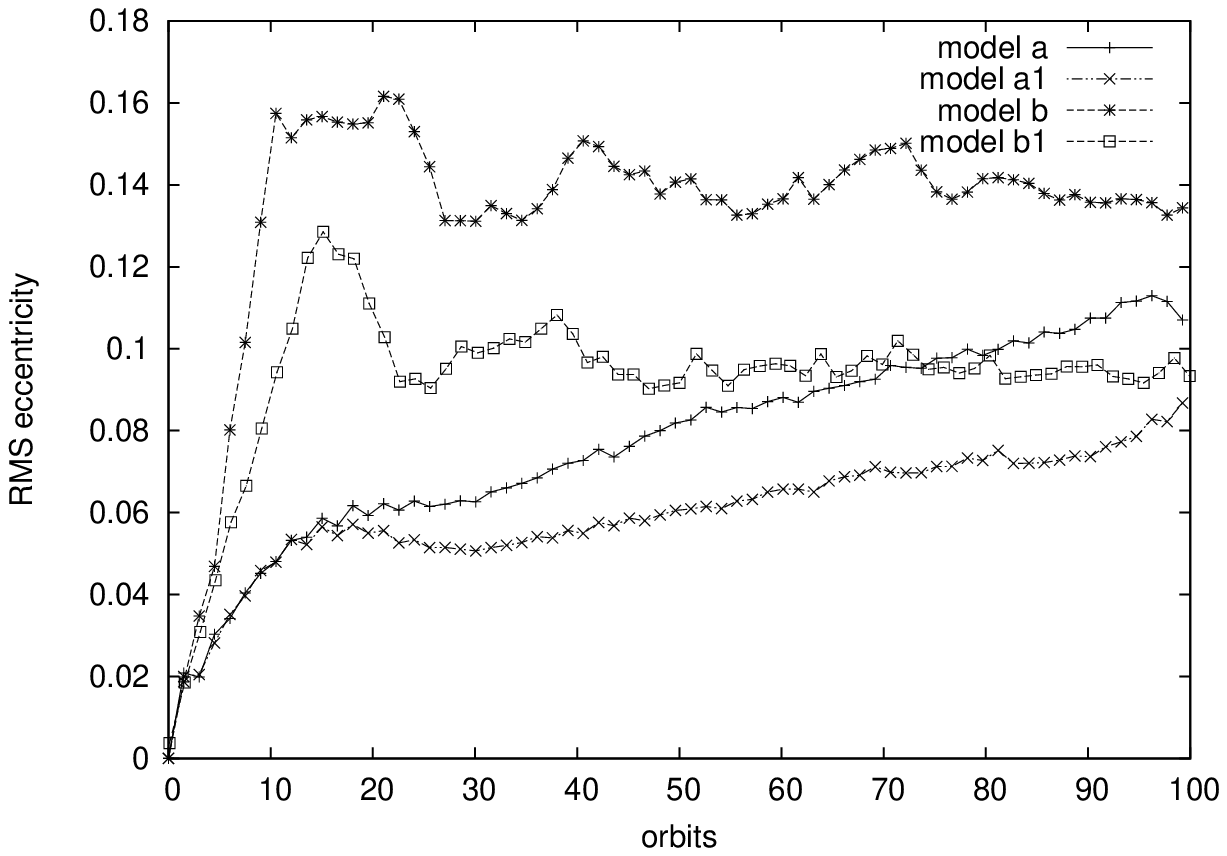}
                       \end{tabular} }
 \caption{Evolution of protoplanet mean (upper plot) and root mean square eccentricity (lower plot) for all four models. Highest values in both plots are reached by model b, whereas model b1 shows a similar behaviour as model a, and the lowest values are reached by model a1.}
   \label{figst3200}%
\end{figure}
The eccentricity of the particles undergoes strong variations in model b1, reaching values of up to $\sim$0.9 in the first 1000 years of the simulation. For model a1 the evolution is proceeding more slowly, but on average stays below 0.4. To compare the outcome for the four models quantitatively we show the time evolution of the protoplanets' mean eccentricity  $\left< e \right>$ as well as their root mean square (rms) eccentricity in Figure \ref{figst3200}. For models a1 and a2 $\left< e \right>$ increases almost linearly from 2000 years onwards, whereas for model b1 $\left< e \right>$ rises quickly to 0.3 and oscillates finally between 0.2 and 0.25. For model b2 from 10 orbital periods on we find a linear increase in $\left< e \right>$ with superposed oscillations. From 70 to 100 orbits the evolution of $\left< e \right>$ is lower than that of model a1. To give an appropriate picture of the state of the system we also computed the rms eccentricity (lower plot of Figure \ref{figst3200}), which is of the order of $\left< e \right>$, reflecting the fact that due to the gravitational perturbation of the secondary we find a strong variation of the particle eccentricity within the computational domain. 
The RMS values of the eccentricities for models a1 and a2  increase linearly from $\sim$ 15 orbits onwards, reaching values  between 0.15 and 0.1 (model a1) and 0.8 and 0.9 (model a2) in the last ten orbits of the simulation. Model b1, in contrast, shows a general increase in the protoplanets' eccentricities which reaches a value of 0.16 within the first ten orbits but slightly decreases to 0.135  at the end of the run. In model b2 the RMS rises to 0.13  within 20 orbits and decreases to values below 0.1. It again is lower than values of model a1 in the range of 70 to 100 orbits. This shows that the influence of a dynamically evolving gas disc onto the protoplanets causes an increase in particle eccentricities, whereas a quiet disc damps particle eccentricity even when we solely take disc gravity into account.

\begin{figure*}
 \resizebox{\hsize}{!}{\begin{tabular}{lll}
                          \includegraphics{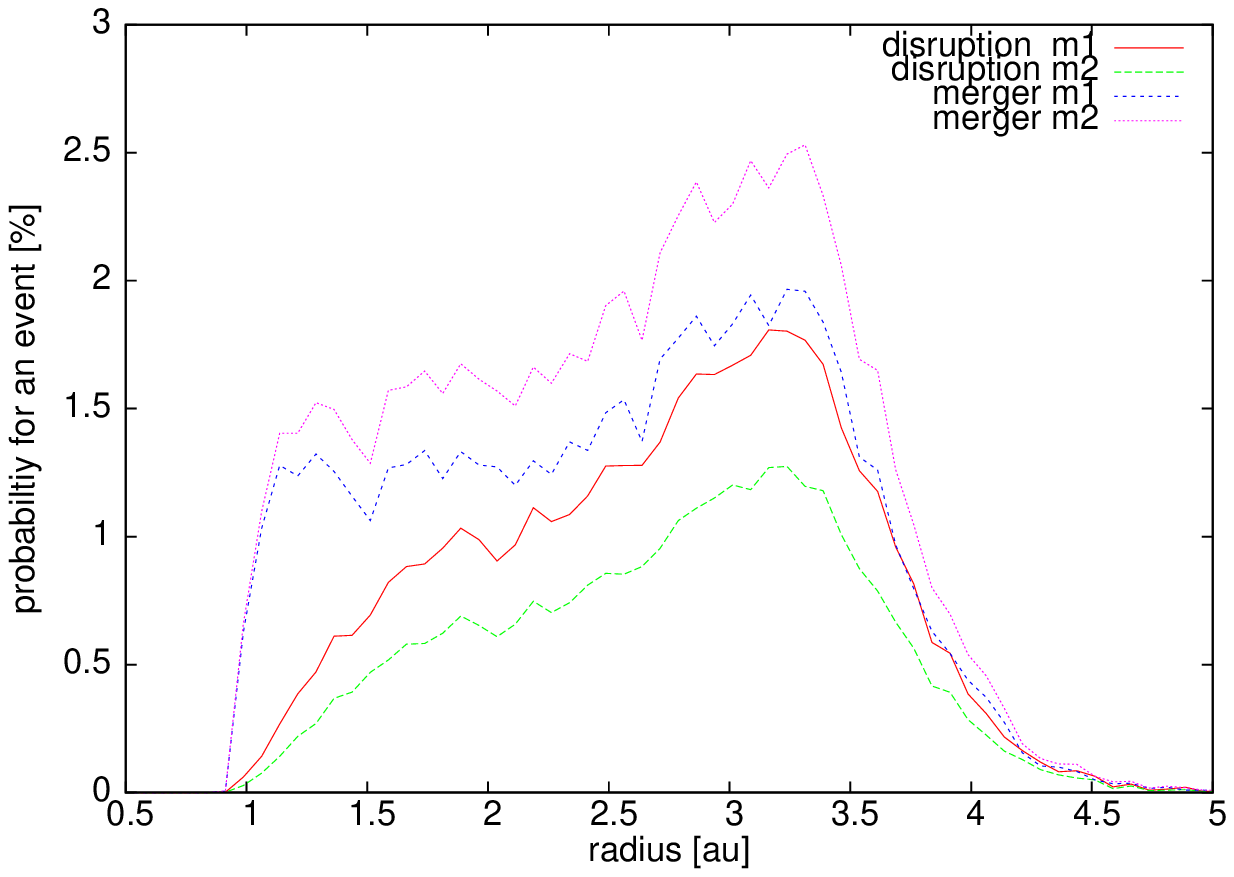}&
                          \includegraphics{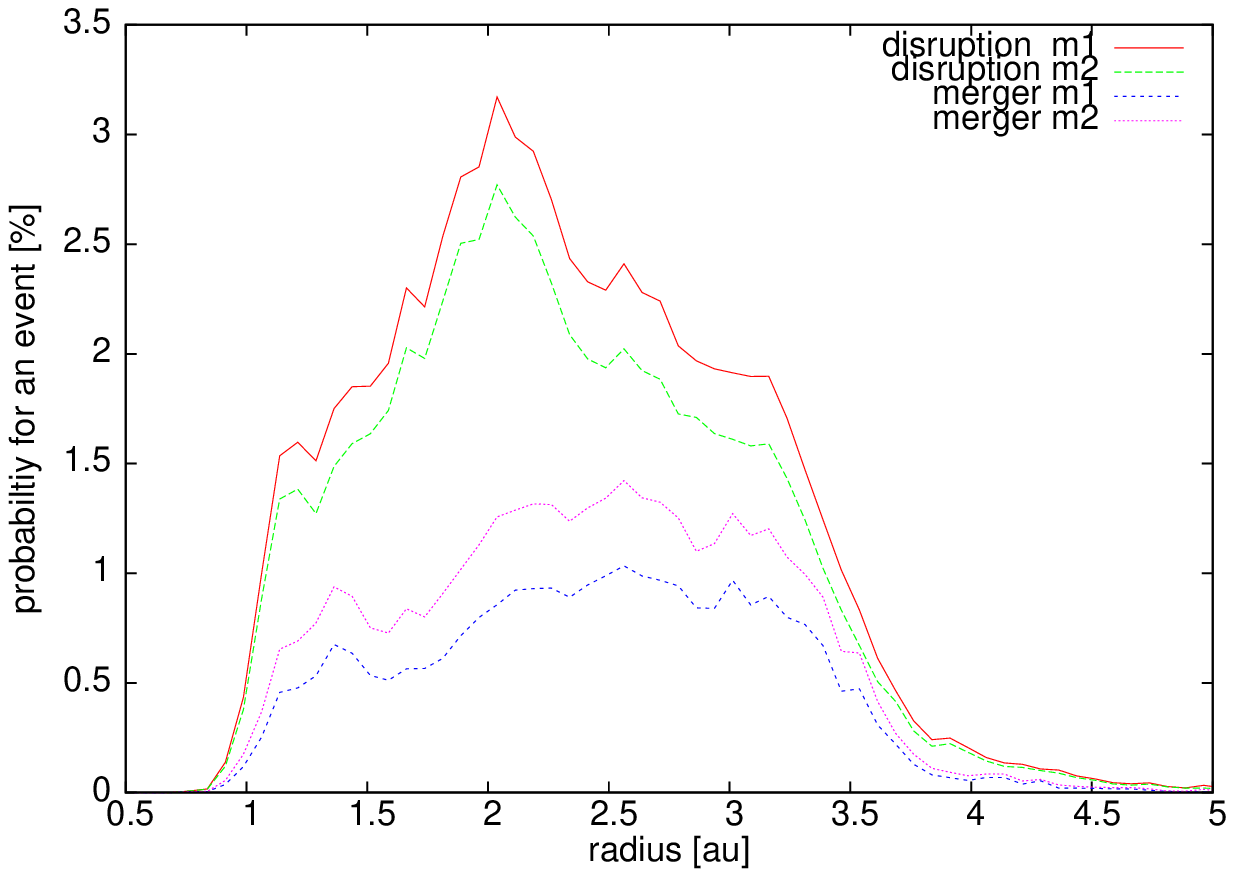}&
                          \includegraphics{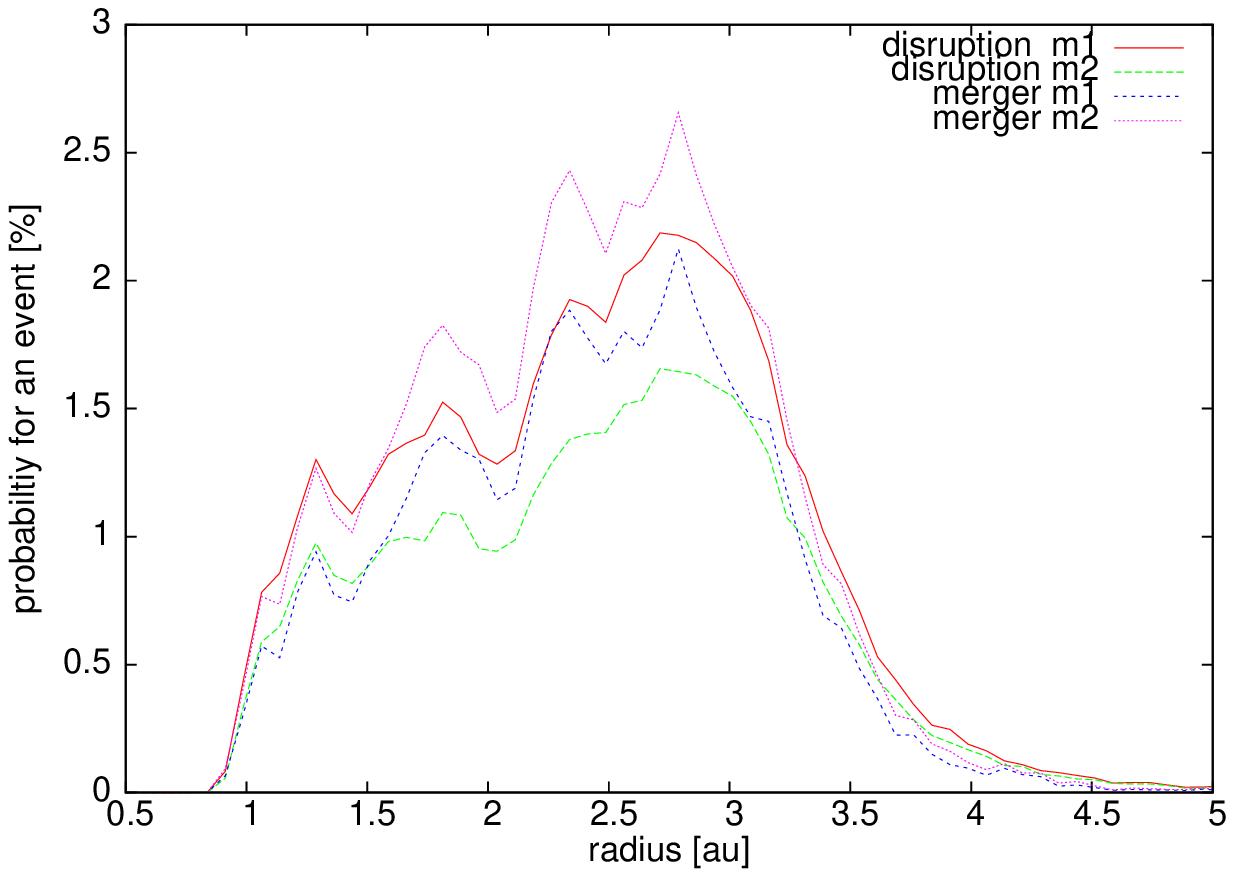} \\
                          \includegraphics{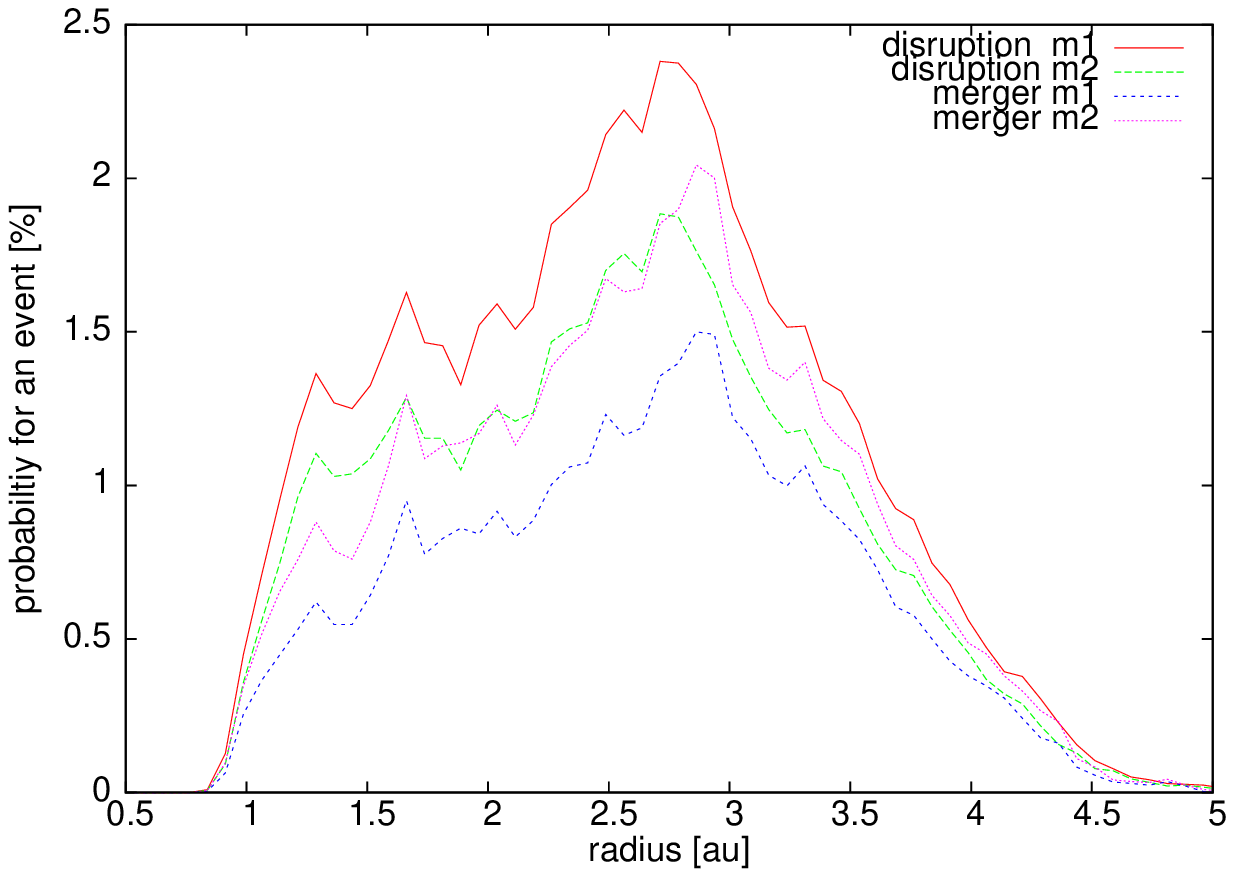} &
                          \includegraphics{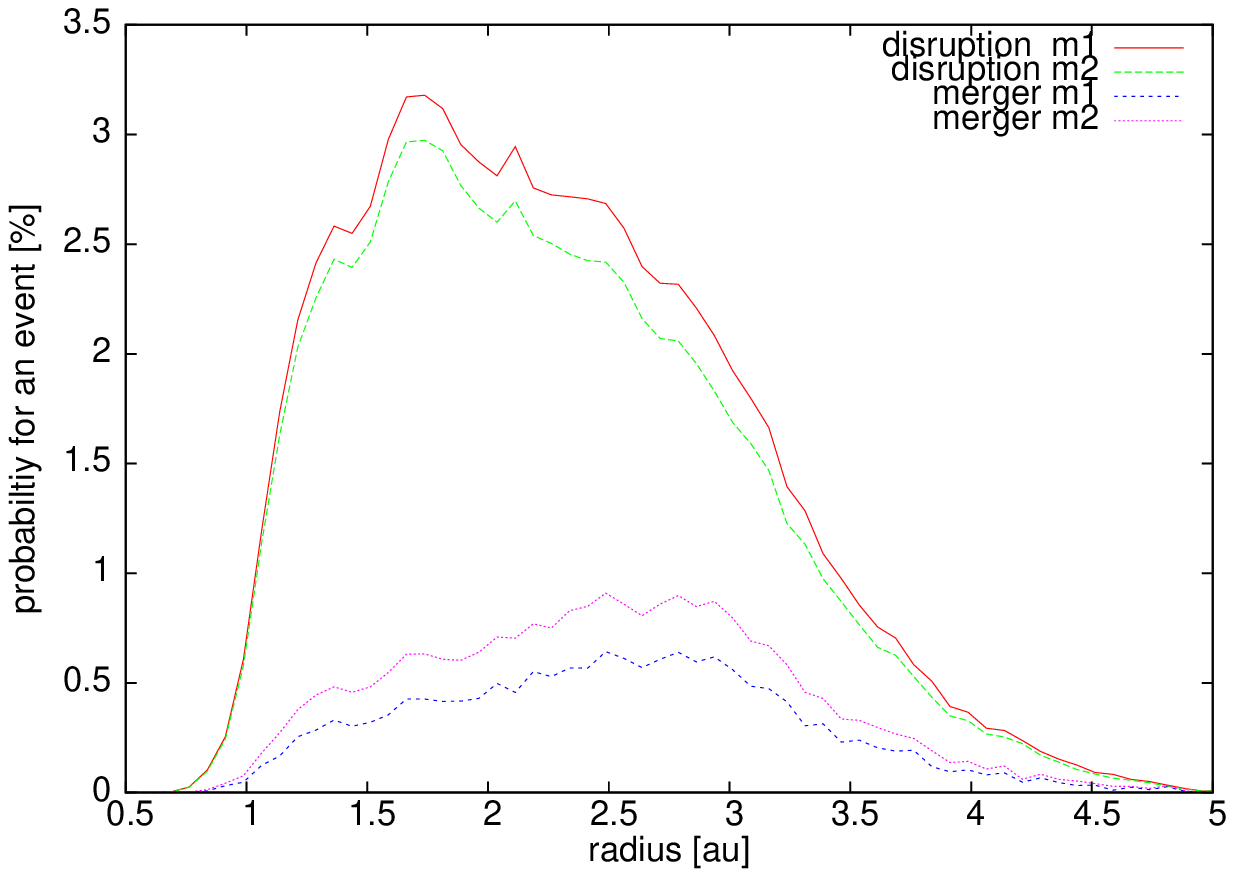} &
                          \includegraphics{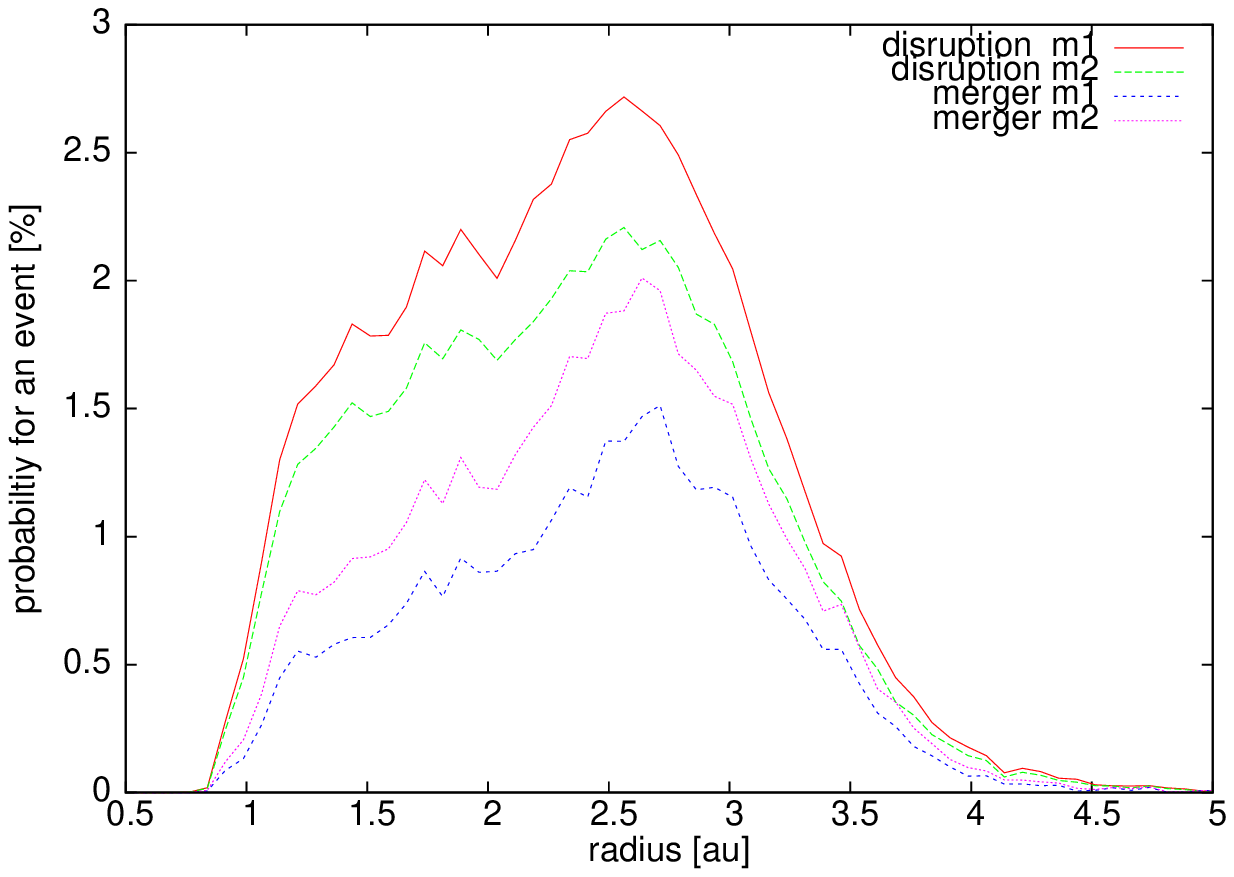}  
                       \end{tabular} }
 \caption{Collision probabilities for 30 - 40 orbital periods for model a1 (\textit{upper left}), model b1 (\textit{upper middle}), and model b2 (\textit{upper right}) as well as for 50 - 60 orbits for model a1 (\textit{lower left}), model b1 (\textit{lower right}), and model b2 (\textit{lower right}). We show the plot radius (in au) versus the probability for an event (in percent) for disruption and merging for methods m1 and m2 (explained in the text).}
   \label{figsmerger}%
\end{figure*}

For the case of a gas-free disc we find that our simulation does not show the wavy forced eccentricity patterns encountered in other studies where the self-gravity of the protoplanets has been neglected \citep{2000ApJ...543..328M,2006Icar..183..193T,2008MNRAS.386..973P}. Such patterns are quickly removed by orbital crossings that start at the outer edge of the system and propagate inside. Because none of the orbit crossing protoplanets was purged from the simulations, the number of potentially high velocity collisions is most likely overestimated and the vanishing intermittent phasing as well the lack of the wavy forced eccentricity patterns may be nothing else than a numerical effect of our applied conditions. \\
In a second step we performed simulations where we computed differential velocities ($dv$) for each close encounter of embryos. A close encounter occurs when an embryo enters the Hill sphere of another. For numerical reasons we computed this in two time intervals: (1) from the 30th to 40th binary orbit (I (30-40)) and (2) from the 50th to 60th binary orbit (I (50-60)), and recorded all encounter events together with the velocities and positions of the involved bodies. The two intervals were chosen for two reasons: (i) as mentioned before, it needs time for perturbations induced by the secondary to move inward and distort the innermost ring (in the worst case this needs $\sim$ 2000 yr, corresponding 30 orbital periods of the binary) and (ii) in all simulations we found periastra alignment for all bodies. For models a1 and a2 this alignment is destroyed very quickly, while for models b1 and b2 in the region of semi-major axes between 1.8 to 3 au this pattern remains for 60 orbits.  This periastra alignment might be indicate an enhanced probability of planet formation since even for highly eccentric bodies such an alignment would lead to low differential velocities and consequently to less violent collisions. \\
 We furthermore computed probability distributions where we divided the radial range (0.5 - 8 au) into 100 rings and counted the number of close encounters within each ring. We distinguished two cases: merging and disruption. A merger occurs if the difference velocity of the two bodies is lower than $v_{esc}$, otherwise it is assumed as a disruption. This is a very simplified criterion that - compared with a more sophisticated model \citep{2012ApJ...745...79L,2012ApJ...751...32S} - overestimates the disruption case. We did not take into account the more elaborate major collision regimes (hit-and-run, disruption, and super-catastrophic disruption) since this is beyond the aim of our study. This means in our case, for instance, that what we call a disruptive event does not necessarily lead to destruction of the bodies, they might also just undergo a hit-and-run scenario, which we cannot properly resolve. The application of these regimes will be carried out in a forthcoming article. \\ 
 For now we wish to show that different types of disc behaviour (dynamically evolving or quiet) can influence protoplanetary orbits such that the probability of further growth may be reduced.  In our study the escape velocity of one body is given by $v_{esc}=\sqrt{2GM_p/R_p} \thicksim $ 1.957 km/s (method m1) or for two bodies by $v_{esc}=\sqrt{4GM_p/R_p} \thicksim $ 2.768 km/s (method m2), where $M_p$ and $R_p$ are the mass and radius of a body. Method m2 is definitely the upper boundary of merging since we assumed that all material has agglomerated into one body. We considered m1 to be a realistic intermediate method as a second indicator to monitor what the common probability for merging or disruption is. Applying method m1, we monitored if one of the bodies could leave the gravitational sphere of influence of the second one, while in m2 we are interested to see whether the body can leave the sphere of influence of the sum of two masses.  From this it follows that a disruptive impact occurs when the differential velocity is higher than $v_{esc}$. \\ 
The probability distributions where computed such that the numbers of mergers and disruptive impacts within each bin were divided by the total number of collisions for each time interval I (30-40) or I (50-60). The result is shown in Figure. \ref{figsmerger}, where plots for model a1 (upper/lower left), model b1 (upper/lower middle), and model b2 (upper/lower right) are shown. The upper panels display the probability distributions for I (30-40) and the lower plots for I (50-60). For model a1 the probability for a merger in I (30-40) is in both cases (m1 and m2) always higher than for disruption and peaks around 3 to 3.5 au. In I (50-60) disruption overwhelms merging in case of m1, but for m2 the distributions for merging and disruption are almost equal and peak around 2.5 to 3 au. For model b1 the  situation is different. There the probabilities for disruptions are two to three times higher than for a merger for both methods (m1, m2) and both computational intervals, and they peak around 2 au for disruption and 2.5 au for merger. The distributions for model b2 are almost similar to those for a1. In the first time interval I (30-40) for case of m1 merger and disruption probabilities are similar, where disruption slightly outbalances merger. For m2 plots for merger and disruption are clearly distinct and merger dominates disruption. The distributions peak around 2.5 to 3 au. In the second interval I (50-60) for m1 the distributions for merger and disruption are clearly separated and disruption dominates by a factor of two; for m2 they are similar and the disruption distribution is slightly higher than that of merger. The distributions peak around 2.5 au. \\
For models a1 and b1 the distributions show different peaks. For a disruption event chances are higher in the inner part of the disc (1 - 2.5 au), while for a merger the outer part (2 - 3 au) is more advantageous. Comparing models a1 and b1, one clearly sees that a dynamically evolving gas disc could alter the planet evolution tremendously. This accretion-hostile environment can be explained by the action of periodically occuring and inward-moving spiral waves in the disc that are exited by the secondary together with periodically varying disc eccentricity. This leads to excitations of the embryos and thus higher encounter velocities. But as mentioned before for model b1, we also find that the peak probabilities for merger and disruption are not aligned, giving a small chance for an embryo to grow in the region between 2 - 3 au. Comparing models a1 and b2, we see almost similar distributions but with a peak shifted  towards the star for model b2. For b2 the influence of the disc is visible through slightly elevated disruption distributions, whereas for a merger they did not change. It seems that a cold disc does not hinder planet growth strongly. \\

For models a1, a2, b1, and b2 we implicitly assumed that the planetesimal accretion phase was successful and only investigated the later stages of planet-formation. This condition is far from being granted. Furthermore one has to keep in mind that the given initial distribution of protoplanetary masses and orbital parameters is rather artificial and was chosen so that the gravitational interplay between the disc and the protoplanets could be studied on short time-scales, and therefore a detailed study that varies the initial conditions of the particles needs to be conducted.\\

%__________________________________________________________________
%__________________________________________________________________
\section{Summary and conclusions}
 \label{sec:sum_con}
We investigated the evolution of particles embedded in a protostellar disc by using orbital parameters similar to the well-known $\gamma$ Cephei system \citep{2007A&A...462..777N}. Introducing our hydrodynamical 2D code, which is suitable to compute several thousands of self-gravitating particles that interact gravitationally with a viscous and radiative circumprimary disc within a binary star system, we presented a series of tests that validated our program (see the appendix \ref{sec:test}). To our knowledge, this program is the only one at the moment that can handle this many particles and calculates their influence on each other and on the disc. By simulating a coplanar binary-disc system with a grid centred on the primary, we compared the evolution of two distinct models of disc-protoplanet interactions with a purely hydrodynamical reference model. Unlike in similar attempts, such as \citet{2012A&A...539A..18M}, the mass loss of the disc during our simulations is small. This is due to the more carefully chosen initial density truncation as well as the sub-Keplerian initial azimuthal velocity.\\
As expected from the total particle-to-gas-mass ratio of $10^{-2}$, the disc evolution is not strongly affected by the particles. But a phase drift in the disc's mass-weighted eccentricity and argument of pericenter occurs between solutions where the protoplanets' influence on the disc is either incorporated or neglected.\\
 Furthermore, we showed that \nbody $ $ relaxation processes occur faster in presence of an dynamically evolving and interacting gas disc, with average protoplanetary eccentricities almost twice as high as in non-interaction models. We also found that the relaxation time depends on the smoothing parameter, which introduces a numerical viscosity and keeps eccentricities on low values. The growth from embryos to planets within a dynamically evolving gas disc is strongly altered by a dynamically evolving disc, which leads to a decreased probability for planet evolution a least in the inner parts of the gas disc. For a cold disc we found distributions almost similar to a solely \nbody $ $ case, but with slightly elevated disruption distributions, which favours them for planet formation. Reasons for these differences between the models are numerous. The influence of the gas and chaotic energy transfer due to close encounters certainly play a role, but more studies are required to identify all contributors. For a gas-free disc we found that our simulation does not show the wavy forced eccentricity patterns encountered in other studies, where the self-gravity of the protoplanets has been neglected \citep{2000ApJ...543..328M,2006Icar..183..193T,2008MNRAS.386..973P}. Such patterns are quickly removed by orbital crossings starting at the outer edge of the system and propagating inside. Since none of the orbit-crossing protoplanets was purged from the simulations, the number of potential high-velocity collisions is most likely overestimated and the vanishing intermittent phasing as well the lack of the wavy forced eccentricity patterns may be nothing but a numerical effect of our applied conditions. There are, of course, several other computational parameters that affect the outcome of the simulations as well, such as the smoothing parameter $\epsilon$ regarding the disc-protoplanet and protoplanet-protoplanet interaction, the type of boundaries (reflecting, outflow, non-reflecting) of the grid in the hydrodynamical part, different flux-limiter functions in the advection part of the code, and the orbit evolution of the secondary star. Future studies will investigate their respective influence and shed more light on the problems of planet formation in binary star systems. 

%__________________________________________________________________
%__________________________________________________________________
\begin{acknowledgements}
      This work was supported by the FWF Projects P-20216-N16 and S11608-N16. M. Gyergyovits is very grateful to W. Kley and T. M\"{u}ller for valuable discussions.
\end{acknowledgements}
%__________________________________________________________________
%__________________________________________________________________
\appendix
\section{standard test for the code}
 \label{sec:test}
 
To validate our code we ran several test cases proposed in literature. The following subsections briefly introduce the tests performed to monitor the respective influence of the viscosity tensor (\ref{subsub:testvisctens}), a gravitational perturber (\ref{subsub:eucomp}) as well as energy loss via irradiation (\ref{subsub:raddisc}) onto the evolution of the circumprimary disc.

%__________________________________________________________________
\subsection{Viscosity tensor}
 \label{subsub:testvisctens}
 
Because our code applies a viscosity tensor, we used the  method reported by \citet{1999MNRAS.303..696K} to test it. We computed the axisymmetric problem of an expanding ring whose initial surface density profile corresponds to a $\delta$-distribution spreading under the influence of a low viscosity and negligible pressure. Assuming Keplerian rotation, the analytical solution of the surface density $\Sigma(r)$ can be given in terms of Bessel functions:
 
\begin{equation}
 \Sigma \left(  r \right) = \frac{C}{\tau x^{1/4}} \exp{\left( -\frac{1 + x^2}{\tau} \right)} I_{1/4} \left( \frac{2x}{\tau} \right).
\end{equation}

\begin{figure}
 \resizebox{\hsize}{!}{\includegraphics{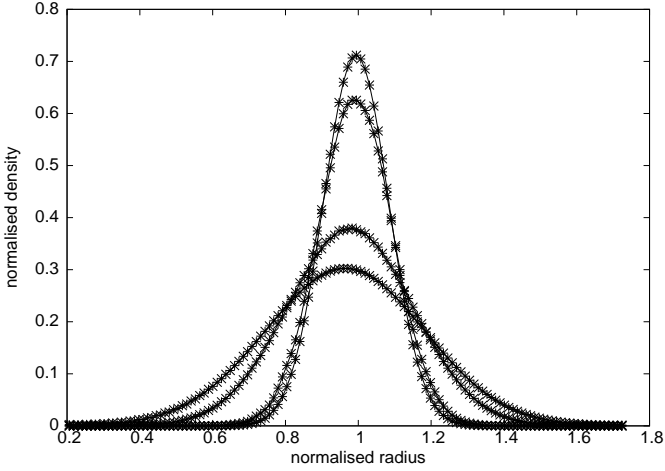}}
 \caption{Normalised surface density for $\tau = 0.016$, $\tau = 0.021$, $\tau = 0.058$,
          $\tau = 0.091$. The solid lines indicate the analytical solution.  
          The initial profile was set according to the analytical solution at $\tau = 0.016$. For details see section \ref{subsub:testvisctens}.}
   \label{figtestvisctens}%
\end{figure}

Here, $x$ denotes the normalised radius $x = r/r_0$ where the initial ring was located at $r_0$, $C=1/(\pi r^2_0)$ and $I_{1/4}$ is the modified Bessel function. The dimensionless time is denoted by $\tau = t/t_v$ given in units of the viscous spreading time $t_v = r^2_0/(12 \nu)$. We computed the full equations of motion on a 128 $\times$ 128 grid using an isothermal approach where the sound speed is set to zero. Figure \ref{figtestvisctens} shows the evolution of the ring for different dimensionless times. Initially, the ring is located at $r_0=1$. The starting time $\tau=0.016$ was chosen to be consistent with the initial conditions of \citet{1999MNRAS.303..696K}, so that the initial distribution no longer resembles a perfect delta distribution. Instead, the ring has already spread somewhat. Our constant dimensionless viscosity was chosen to be $\nu=10^{-6}$ and no artificial viscosity was introduced. The numerical solution (points) agrees well with the analytical results (lines).

%__________________________________________________________________
\subsection{Disc - planet interaction}
 \label{subsub:eucomp}

 The disc-planet interaction was tested via recomputing the case of Jupiter in a viscous and a non-viscous disc, as can be found in \citet{2006MNRAS.370..529D}. We also applied an additional artificial viscosity of von Neumann \& Richtmyer type \citep{1992ApJS...80..753S}, where $C^2 = 2$. According to the test conditions given in this article, we used $n_r \times n_{\phi}=(128,384)$ cells in radial and azimuthal direction, reflecting boundaries, and applied a damping function 
 
 \begin{equation}
  \frac{dx}{dt} = - \frac{x-x_0}{\tau}R(r),
  \label{equdamp}
 \end{equation}
  
where $x$ represents the surface density and velocity components, $\tau$ is the orbital period at the corresponding boundary, and $R(r)$ is a parabolic function \citep[equation 10]{2006MNRAS.370..529D}. The semi-major axis of the planet was set to $a = 1$ au and the planet-to-star-mass ratio $\mu$ was 0.001. The initial surface density is constant:
 \begin{equation}
  \Sigma_0=0.002\frac{M_s}{\mathrm{\pi} a^2}.
 \end{equation}
 This entails a total disc mass of $M_\mathrm{d}=0.012 \ \mathrm{M_{\sun}}$. $M_s$ is the mass of the star. We used a locally isothermal approach and $H/r = 0.05$, which results in $c_\mathrm{s}=0.05 v_\mathrm{K}$ where $c_\mathrm{s}$ is the isothermal sound speed and $v_\mathrm{K}$ is the local Keplerian velocity. The grid extends from 0.4 to 2.5 au, and we applied the artificial damping of Eq. (\ref{equdamp}) between 0.4 and 0.5 as well as between 2.1 and 2.5 au. The planet mass was slowly increased to its final value during its first five orbits according to 
 \begin{equation}
  M(t)=M_\mathrm{p}\sin^2 \left( \frac{\pi t}{10 P_\mathrm{p}} \right), \qquad t= 0  $ $$to$$ $  5 P_p,
 \end{equation}
where  $P_\mathrm{p}$ is the period of the planet. The kinematic viscosity was chosen to be $\nu = 1 \times 10^{-5}$ in dimensionless units, where $a = 1$, $P_p = 2\pi$ and $M_s = 1 - \mu$. 
We can reproduce the normalised azimuthally averaged surface density after 100 planetary orbits for the non-viscous case and viscous case and also the evolution of the total torque for the whole computation time of 200 orbits shown in \citet{2006MNRAS.370..529D}. 
Averaging the torques between 175-200 periods in dimensionless units produces values of $-7.73 \times 10^{-5}$ for the viscous case  and $-1.75 \times 10^{-5}$ for the inviscid case. The time is given in orbital periods of the planet. Compared with the values of Tables 5 and 6 in \citet{2006MNRAS.370..529D}, our results are similar to the outcome of RH2D,  a code written by \citet{1989A&A...208...98K}. 
%__________________________________________________________________ 
\subsection{Radiative disc}
 \label{subsub:raddisc}

\begin{figure}
 \resizebox{\hsize}{!}{\begin{tabular}{l}
                          \includegraphics{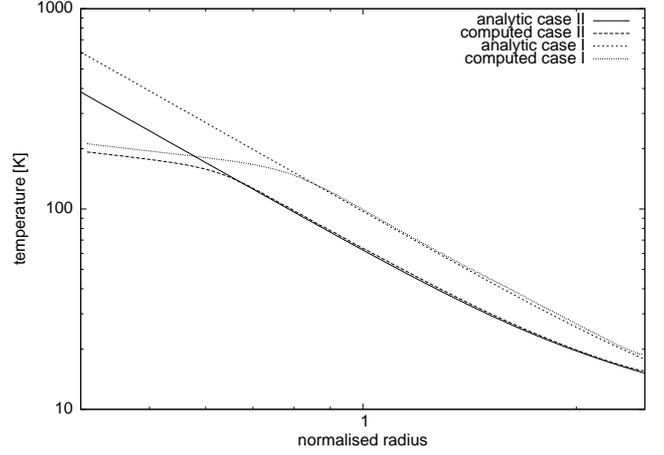}
                       \end{tabular} }
 \caption{Comparison of analytical and simulated disc temperatures for cases I and II. The radius is normalised $r = a/[5.2 au]$ and the temperature is given in Kelvin. }
   \label{figtemp}%
\end{figure}

\begin{figure}
 \resizebox{\hsize}{!}{\begin{tabular}{l}
                          \includegraphics{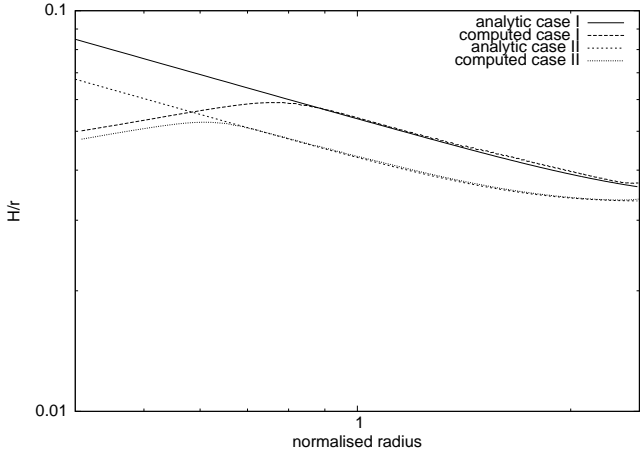}
                       \end{tabular} }
 \caption{Same as Figure ~\ref{figtemp}, but for aspect ratios $H/r$.}
   \label{figscaleheight}%
\end{figure}

\begin{figure}
 \resizebox{\hsize}{!}{\includegraphics{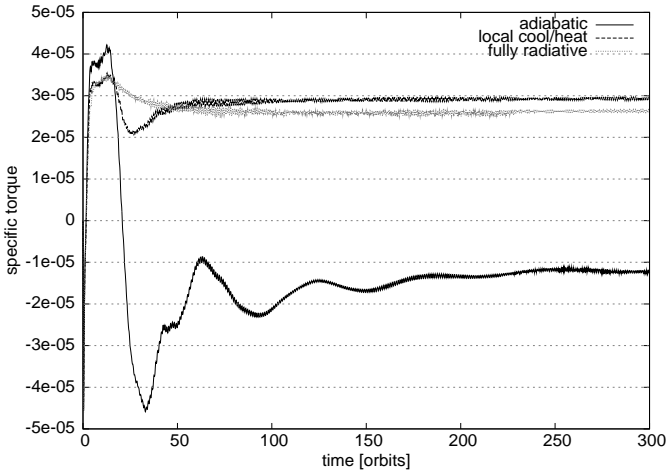}}
 \caption{Time evolution of the specific total torque T [$(a^2 \Omega^2)_{Jup}$] exerted by 
the disc on a planet of 20 $M_{Earth}$ for the adiabatic, local cool/heat, and fully radiative case. }
   \label{figskctorques}%
\end{figure}
 
To test our implementation of the energy equation we compared our code with the fully radiative reference model by \citet{2008A&A...487L...9K} (see also Eq. \ref{enquation} of this article). The protoplanetary disc is treated as a two-dimensional, non-self-gravitating gas that can be described by the Navier-Stokes equations. The embedded planet is modelled as a point mass that orbits the central star on a fixed, circular orbit. For the planetary potential, we used a smoothing length of = 0.6H, where H is the vertical scale height of the disc. To calculate the gravitational torques acting on the planet, we  excluded the inner parts of the Hill sphere of the planet.
 The test was carried out twice. 
\begin{description}
\item[\textbf{Case I:}] in this case we implemented the energy equation given in \citet{2008A&A...487L...9K} and also used the adiabatic sound speed. 
\item[\textbf{Case II:}] for the second case the method suggested in \citet{2012A&A...539A..18M} was implemented using only  the isothermal sound speed. 
\end{description}
The numerical parameters for both cases were chosen as follows: a two-dimensional (r - $\phi$) computational domain consisting of a ring-shaped protoplanetary disc, which is centred on the star. This domain extends from $r_{\mathrm{min}}$ = 0.4 $a_{\mathrm{Jup}}$ to $r_{\mathrm{max}}$ = 2.5 $a_{\mathrm{Jup}}$, where $a_{\mathrm{Jup}} = 5.2$ au is the semi-major axis of Jupiter. The total disc mass was set to $M_d=0.01 \ \mathrm{M_{\sun}}$ and the constant kinematic viscosity $\nu = 10^{15} \ \mathrm{cm^2/s}$. Following a power law, the initial density distribution of the gas is given by $\Sigma(r) = \Sigma_0 r^{-1/2}$, the temperature $T(r) \propto r^{-1}$ and the rotation is sub-Keplerian. A simulation pre-run of 200 orbits of Jupiter was performed to ensure that the disc is in equilibrium for the actual test calculations. The disc boundary conditions were chosen to be reflecting, and dampening was applied in the regions [0.4$\ a_{\mathrm{Jup}}$, 0.5$\ a_{\mathrm{Jup}}$] and [2.1$\ a_{\mathrm{Jup}}$, 2.5$\ a_{\mathrm{Jup}}$]. \\ 
The analytical temperature profile ranging between 0 K and 170 K was derived from the equilibrium of heating and cooling \mbox{($Q_{+}$ = $Q_{-}$)} as well as from fact that the Rosseland mean opacity is $\kappa = \kappa_0 T^2$. We furthermore assumed that the radial velocity of the disc is zero. This gives an equation of sixth order in T, which can be reduced to third order and finally solved to arrive at one real solution. In Figure ~\ref{figtemp} we show a comparison of simulated and analytical disc temperatures for cases I and II. The analytical and computed solutions match very well except for high temperatures, where the assumption $\kappa = \kappa_0 T^2$ is no longer valid. Comparing  cases I and II, the temperature profile in the latter case is lower and the point where analytical and numerical solutions branch moves inward. At this point $\kappa$ starts to follow a different power law in T. A similar behaviour can be seen in the aspect ratio (Figure \ref{figscaleheight}). In both cases the aspect ratio does not increase steadily until the inner boundary is reached. It reaches a highest value of 0.059 at 0.77 $a_{\mathrm{Jup}}$ for case I and 0.0526 at 0.6 $a_{\mathrm{Jup}}$ for case II. The values in case I are higher than in case II and the maximum of the numerical solution is shifted to smaller radii for case II compared with case I. This is expected since $c_s \propto \sqrt{T}$ and H/r = $c_{\mathrm{s}}/v_{\mathrm{K}}$. \\
In a second test we recomputed the specific torque the disc exerts on a 20 $M_{\mathrm{Earth}}$ planet using the implementations of case I for three modes, neglecting different terms in the energy equation \citep[Figure 2]{2008A&A...487L...9K}. Omitting all terms except the first one on the right side in equation (\ref{enquation}) gives an adiabatic model, whereas in the cool/heat case we just neglected the last term. In the fully radiative case we included all terms in the equation. In Figure~\ref{figskctorques} we show that we can find an evolution of the specific torque the disc exerts on the planet, which is similar to the outcome of \citet[Figure 2]{2008A&A...487L...9K}, especially during the time between 200 and 300 planetary orbits. The local cool/heat case matches very well, whereas the specific torque in the fully radiative case is slightly higher than in \citet[]{2008A&A...487L...9K}. The adiabatic case shows stronger oscillations in the beginning and approaches a higher equilibrium value than presented in \citet{2008A&A...487L...9K}. These small differences might arise because we used sub-Keplerian azimuthal velocities for the disc and damped towards these values instead of purely Keplerian ones.

%__________________________________________________________________
%__________________________________________________________________

\bibliographystyle{aa}
\bibliography{paper}

\end{document}